# Estimation of π-π Electronic Couplings

# from Current Measurements

*J. Trasobares[a,λ,δ], J. Rech[b,λ], T. Jonckheere[b], T. Martin[b], O. Aleveque[c], E. Levillain[c], V. Diez-Cabanes[d], Y. Olivier[d], J. Cornil[d], J.P. Nys[a], R. Sivakumarasamy[a], K. Smaali[a], P. Leclere[d], A. Fujiwara[e], D. Théron[a], D. Vuillaume[a] & N. Clément[a,e*]*

[a] Institute of Electronics, Microelectronics and Nanotechnology, CNRS, Univ. of Lille, Avenue Poincaré, BP60069, 59652, Villeneuve d'Ascq France

[b] Aix Marseille Univ., Universite de Toulon, CNRS, CPT, 163 Avenue de Luminy, 13288 Marseille cedex 9, France

[c] Université d'Angers, CNRS UMR 6200, Laboratoire MOLTECH-Anjou, 2 bd Lavoisier, 49045 Angers cedex, France

[d] Laboratory for Chemistry of Novel Materials, University of Mons, Place du Parc 20, B-7000 Mons, Belgium.

[e] NTT Basic Research Laboratories, 3-1, Morinosato Wakamiya, Atsugi-shi, kanagawa, 243-0198, Japan

[λ] These authors contributed equally to the work.

δ: Present address: Department of Chemistry, NUS 3 Science Drive 3 Singapore 117543

The $π$-$π$ interactions between organic molecules are among the most important parameters for optimizing the transport and optical properties of organic transistors, light-emitting diodes, and (bio-) molecular devices. Despite substantial theoretical progress, direct experimental measurement of the $π$-$π$ electronic coupling energy parameter $t$ has remained an old challenge due to molecular structural variability and the large number of parameters that affect the charge transport. Here, we propose a study of $π$-$π$ interactions from electrochemical and current measurements on a large array of ferrocene-thiolated gold nanocrystals. We confirm the theoretical prediction that $t$ can be assessed from a statistical analysis of current histograms. The extracted value of $t$ ≈35 meV is in the expected range based on our density functional theory analysis. Furthermore, the $t$ distribution is not necessarily Gaussian and could be used as an ultrasensitive technique to assess intermolecular distance fluctuation at the sub-angström level. The present work establishes a direct bridge between quantum chemistry, electrochemistry, organic electronics, and mesoscopic physics, all of which were used to discuss results and perspectives in a quantitative manner.

**Introduction**

Interactions between π-systems [1,2] are involved in diverse and important phenomena, such as the stabilization of the double helical structure of DNA,[3] protein folding,[4] molecular recognition,[5] drug design,[6] and crystal engineering.[7] These interactions are of fundamental technological importance for the development of organic-based devices,[8] in particular for organic light-emitting diodes,[9] field-effect transistors[10], or (bio-) molecular devices.[11-16] A key parameter in these interactions is the transfer integral (or electronic coupling energy) parameter $t$, which is included as $t^2$ in simple semiclassical formulations of charge carrier mobility.[17] In symmetric dimers, $t$ is directly related to energy-level splitting of the highest occupied/lowest unoccupied molecular orbital (HOMO/LUMO) due to intermolecular interactions for hole and electron transport, respectively.[8]

The parameter $t$ has mainly been discussed by using photoelectron spectroscopy and quantum-chemical calculations.[18-20] In the ideal scenario for (opto-)electronic applications, $t$ should be deduced directly from electronic measurements in a device configuration and related to the molecular structure. Such knowledge of $t$ would help us to understand and optimize charge transport through molecular systems. For example, cooperative effects, induced by molecule-molecule and molecule/electrode electronic couplings, are attracting substantial theoretical attention.[21,22] The distribution or fluctuation of $t$ plays a key role in the charge transport through organic semiconductors or biomolecules by inducing charge localization or conformational gating effects.[23-25] A Gaussian distribution of $t$ with a standard deviation (SD) in the range of the mean $t$ is usually assumed from thermal molecular motions,[25] but remains to be confirmed experimentally. The experimental measurement of $t$ could potentially be used as an ultrasensitive chemical characterization technique because $t$ is expected to be more sensitive to molecular structural order than other physical constants such as π-π electrostatic interactions (*φ*) measured by Cyclic Voltammetry (CV) (Figure 1a). However, recent efforts to establish correlations between electrochemical and molecular electronics results[26-31] have neglected π-π intermolecular interactions.

To reach these goals, two main issues need to be addressed. A first issue is related to disorder. Structural variability makes it difficult to extract $t$ from electronic measurements because $t$ is extremely sensitive to order at the angstrom level[8]. One recently implemented and elegant way to measure charge transport at the local scale is through photoinduced time-resolved microwave conductivity (TRMC),[32] but this contactless approach differs from the measurement of charge transport in a device configuration. The alternative approach is to reduce electrodes and organic layer dimensions.

A second issue is that comparisons between experimental and theoretical charge transport data are usually qualitative. Even without molecular organization disorder, many parameters influence the measured current including molecule/molecule or molecule/electrode coupling and electron-vibration (phonon) interactions[8,33]. A recent theoretical proposal suggested additional degrees of freedom. Reuter et al. found that quantitative information on cooperative effects may be assessed by statistical analysis of conductance traces.[21,34] This approach is based on the Landauer Buttiker Imry formalism that typically is used in mesoscopic physics for the study of electron transport through quantum dots in the coherent regime. The related experimental model system is a single layer of π-conjugated molecules (quantum dots), which is sandwiched between two electrodes. Thousands of molecular junctions are required for statistical analysis. The authors suggested that cooperative effects between molecules should provide asymmetrical conductance histogram spectra (Figure 1b). Histogram fitting may be achieved by considering the mean and SD of molecule site energies ($\varepsilon$, $\delta\varepsilon$), molecule-electrode coupling ($V$, $\delta V$) and transfer integrals ($<t>$, $\delta t$).[21] This fitting differs from the usual experimental log-normal conductance histogram shape (normal distribution when conductance $G$ is plotted in log scale) reported in single molecule-based molecular electronics (Figure 1b) (see Supplementary Note S1 in Supporting Information [SI] for a detailed history of conductance histograms in molecular electronics).[11,35-41]

Here, we explore π-π intermolecular interaction energies from the electrochemical perspective (coupling between charge distributions) and molecular electronic perspective (coupling between orbitals) using a large array of ferrocene (Fc)-thiolated gold nanocrystals. First, we show that the two peaks observed in voltammograms on these systems can be controlled by the nanocrystal diameter. Each peak corresponds to a dense or dilute molecular organization structure located at the top or side facets of the nanocrystals, respectively. Second, the dense molecular organization structure is resolved by Ultra-High-Vacuum Scanning Tunneling Microscopy (UHV-STM). This structure is used as a reference for estimating $t$ from quantum chemical calculations at the Density Functional Theory (DFT) level. Based on current measurement statistics for ~3000 molecular junctions between the top of the nanocrystals and a conducting atomic force microscope (C-AFM) tip, we confirm the theoretical prediction of histograms that shape is affected by cooperative effects.[21] Furthermore, we extend the previously proposed tight-binding formalism to fit the histograms[21,34]. The estimated electronic coupling energy distribution for $t$ is quantitatively compared with quantum-chemical calculations. The $\varphi$ and $t$ obtained from CV traces and current histograms, respectively, are discussed on the basis of intermolecular distance fluctuations. Finally, we highlight the implications and perspectives of this study to molecular electronics, organic electronics and electrochemistry.

**Results**

**Electrochemical characterization of Fc-thiolated gold nanocrystals**

We selected ferrocenylalkylthiol (FcC$_{11}$SH) as an archetype molecule with a π-conjugated head for electrochemistry[29,30,42,43] and molecular electronics.[12,14,29-46] CV is a powerful tool to gain insights into the molecular organization, extract surface coverage $\Gamma$, and evaluate the energy level of the HOMO ($E_{HOMO} \pm \delta E_{HOMO}$). In particular, as different molecular organization structures usually lead to multiple CV peaks,[47,48] the aim of this section is to demonstrate that molecules located at the top of the Fc-thiolated gold nanocrystals correspond to a single CV peak, from which $\varphi$ can be extracted.

Figure 2a-c (and Figure S1,S2 in SI) show the experimental setup with NaClO$_4$ electrolyte (0.1 M) facing Fc molecules and the Au nanocrystal electrodes.[49] We have previously demonstrated the possibility of performing CV on Fc-thiolated gold nanocrystal surfaces,[14] although we studied only one dot diameter and did not investigated cooperative effects. CV cannot be performed at the single-dot level with these molecules because the currents are too weak (≤ fA range).[14,50,51]

First, we assess $E_{HOMO}$ and dot-to-dot dispersion in $E_{HOMO}$ ($\delta E_{HOMO}$) by CV. The voltammogram is averaged over millions of nanocrystals with a few Fc molecules per nanocrystal (diluted in a $C_{12}SH$ matrix) to avoid cooperative effects (Figure 2d). The peak energy position is in the expected range for Fc molecules (0.41 eV vs Ag/AgCl).[12,14,47,42] Furthermore, the voltammogram width at half maximum (FWHM) is close to 90 mV for the main peak, i.e. the theoretical value in the absence of interaction between redox moieties.[27] This result suggests that $\delta E_{HOMO}$ is less than 45 meV (Figure S3).

Figure 2e-g show conventional CV results for nanocrystals (of different diameters) that are fully covered with $FcC_{11}SH$ molecules (raw CV curves are shown in Figures S4-S6). Peak splitting can be observed[42-44]. The peak area is related to the total faradic charge and, therefore, to the number of molecules per dot. Only the number of molecules per nanocrystal related to peak 1 significantly varies with nanocrystal diameter $D$ (Figure 2h). Based on a simple model with a truncated conical shape for dots (Figure 2c), we suggest that peak 1 corresponds to molecules at the top of the dot, whereas peak 2 corresponds to molecules on the side of the dot (Figure 2h, inset). Thus, the density of molecules is smaller on the sides ($\Gamma$~2 nm²/molecule) than on the top ($\Gamma$~0.39 nm²/molecule) of the nanocrystals (see Figure 2h for fits and Methods for details). In other words, a highly ordered structure corresponding to a single peak in the voltammogram can be successfully formed on the top of the gold nanocrystals. This hypothesis is consistent with FWHM ≥90 mV for peak 1 (global repulsion between Fc

moieties in the electrolytic media used[56]) and FWHM ≤90 mV for peak 2. The position and shape of the second peak can be explained by a local change of the environment (presence of $Na^+$ counter ions from negatively charged silica at the dot borders and pH>2; Figure 2i) and a modification of ion-pairing equilibrium[47,52] (fewer $ClO_4^-$ ions at dot borders due to $SiO^-$ surface sites). CV on the smallest dots results in a single peak whose width is smaller than the width expected at room temperature without molecular interactions. This result could be technologically useful for improving the sensitivity of electrochemical biosensors beyond the thermal Nernst limit.[53-55]

The strength of electrostatic interactions for molecules located at the top of the gold nanocrystals can be quantitatively assessed by the extended Laviron model[27,56,57] (see Supplementary Methods). Coulomb interactions ($\varphi$ when Fc moieties are fully oxidized) tune the FWHMs of the voltammograms because they are modulated by the fraction of oxidized species. Reasonable fits can be obtained with $\varphi$ = 4.5 meV for all dot diameters (see Table S1 in SI for fit parameters). The $\varphi$ obtained from CV will be linked to $t$ from the current measurements in the Discussion section.

**Estimation of $t$ from quantum-chemical calculations**

The self-assembled monolayer (SAM) structure on a gold substrate has been resolved by UHV-STM (Figure 3a) and used as a reference for DFT calculations. The STM image shows a regular structure of elongated shapes corresponding to groups. The extracted average area per molecule (0.40 $nm^2$) is in agreement with our CV results and is slightly larger than the 0.36 $nm^2$ considered for a hexagonal structure with a diameter of 0.66 nm per Fc[58,59]. The area corresponds to a configuration in which Fc units are at the same level in the vertical position. Each Fc unit forms a tilt angle of 56° ± 15° with respect to the surface normal (Figure S7), consistent with estimates obtained by obtained by Near-edge X-Ray absorption fine structure spectroscopy (60° ± 5°)[60] and by molecular dynamics simulations (54° ± 22°).[60]

When molecules are organized as in Figure 3b, $t$ can be calculated by DFT for two neighboring Fc units (fragments). This simulation is only based on the Fc units and not on the full $FcC_{11}SH$ molecule because the contribution of the saturated part of the molecule to $t$ is negligible. As structural fluctuations in monolayer organization are expected experimentally, we compute $t_a$ and $t_b$ between fragments of molecules 1 and 3 and molecules 1 and 2 at different positions along the X and Y axes. Figure 3c shows $t_b$ when molecule 2 moves along the X axis in a collinear geometry. $t_b$ strongly depends on displacements of molecule 2 at the angstrom level because $t$ is related to the electronic (rather than spatial) overlaps between $\pi$ orbitals.[61-64] Maxima are in the 20−30 meV range. Figure 3d shows the evolution of $t_b$ as a function of the variation of the intermolecular distance $d$ ($\delta d$) around the equilibrium position, without lateral displacement. The decay ratio $\beta_b$=1.94/ Å is close to the tunnel decay ratio in molecular electronics. Similar results are obtained for $t_a$ (cofacial geometry; see Figure S8). For consistency with our previous studies, the B3LYP functional (see Methods) has been chosen due to the good agreement with mobility values extracted from the TRMC technique.[32] A recent theoretical study illustrated that B3LYP behaves very similarly to long-range corrected functionals and that the size of the basis set has a weak impact on the calculated transfer integrals.[65]

Overall, the results indicate that the π-conjugated Fc molecules are electronically coupled and suggest that a signature of cooperative effects should be observed on current measurements.

**Cooperative effects on current histograms**

We have conducted the statistical study proposed in ref.21 (i.e., current histograms). "Nano-SAMs"(i.e., SAMs with diameters of a few tens of nanometers) are ideal for this experiment. Use of nano-SAMs enables us to obtain sufficient molecules for cooperative effects, but limits the number of molecules to avoid averaging over many molecular structures, grain boundaries, and defects. The C-AFM, as the top electrode,[66] is swept over thousands of nanocrystals.[39] We previously showed that log-normal histograms are systematically obtained when such a statistical study is performed with nano-SAMs composed of alkyl chains without π groups.[39] In contrast, as predicted in ref.21, we find that the presence of cooperativity between π-conjugated orbitals (in the head group) affects the line shape of histograms. Figure 4a is the current histogram obtained on $FcC_{11}SH$ nano-SAMs at -0.6V for 45-nm-diameter gold nanocrystals (2D histogram corresponding to different tip biases is shown in Figures S9). The related histogram line shape can be nicely fitted with asymmetric double sigmoidal function when the current histograms are plotted in log scale (see Methods and Table S2 for fitting parameters). In the case of 15-nm-diameter nanocrystals, a second peak, corresponding to another molecular organization structure[39] appears at a lower current in the histograms (Figure 4b). We suggest that this peak, which is barely seen in the histograms, is averaged on larger dots. Fitting parameters for the main current peak are almost unchanged (see Table S2). When FcC$_{11}$ molecules are diluted 1:1 with dodecanethiol molecules (C$_{12}$SH) to reduce coupling between $\pi$ molecules, the log-normal histogram is recovered (Figure 4c), similarly to alkyl-chain-coated nanocrystals.[39]

We tried to fit the current histograms using a coherent scattering formalism similar to the one proposed in ref.21, with the additional consideration of asymmetrical coupling to the electrodes and the possibility of simulating up to 9×9 molecules (only two molecules were considered in ref.21). Figure 5a illustrates the modeled system. Each Fc molecule is considered as a single-level quantum dot coupled to both electrodes. Dots are coupled together with coupling term *t* in a tight binding model. This coupling term is equivalent to the transfer integral in DFT. For simplicity, *t* is considered to be identical along both axes in the plane. Each molecule within a molecular junction composed of *N*×*N* molecules has the same parameters $\varepsilon$ (molecule orbital energy), $V_t$, $V_b$, (molecules coupling to top and bottom electrodes, respectively) and *t*. Cooperative effects, whose strengths are controlled by parameters $V_t, V_b, t, N$, cause a smearing out of the energy-dependent transmission coefficient shape, with a peak transmission being less than one[21] (see Figures S13-S15 in SI for additional illustrations). The current, obtained from the integral of the transmission coefficient over a range of energy set by the external potential, depends on these parameters accordingly (see SI Methods). To generate current histograms, $V_t, V_b, t, \varepsilon$ are chosen from Gaussian distributions with predefined means and SDs (e.g. eq.1a) for each individual molecular junction. For *t*, we additionally considered eq 1b to explicitly consider the fluctuation of the intermolecular distance (see Figure 3d).

$$t = <t> + \delta t \quad (1a)$$

$$t = t_0 \cdot exp(-\beta \delta d) \quad (1b)$$

, with $\delta d$ in eq 1b is chosen from a Gaussian distribution. A step-by-step fitting protocol is detailed in the Methods section and Figures S10, and S11 in SI. Figure 5b illustrates the possibility of generating histograms that reproduce the experiments. Optimized parameters for 9×9 molecules using eq 1a for $t$ ($t$ = 0.04 eV, $V_t$ = 0.401 eV, $V_b$ = 0.144 eV, $\delta\varepsilon$ = 40 meV, $\delta t$ = 0.14 eV, and $\delta V$ = 22 meV) are in the range of those considered in ref.21 based on ref. 67 where $t$ was 0.1 eV, $V_t = V_l$ = 0.6 eV, $\delta t$ = 75 meV, $\delta V$ = 37 meV, and $\delta\varepsilon$=30 meV). Considering eq 1b for $t$ gives an even better fit to experimental data with $t_0$=0.34 eV, $\beta$ = 1.96/Å, and SD($\delta d$) = 0.8Å. When intermolecular coupling is suppressed ($t$ = 0) while keeping other parameters constant to mimic the diluted monolayer (Figure 5c), the resulting log-normal histogram reproduces the experimental results (Figure 5d).

**Discussion**

In molecular or organic electronics, comparisons of experimental and theoretical charge transport data are usually qualitative. Therefore, any step towards a more quantitative analysis is important to the field.

A strong coupling asymmetry of $\alpha = V_t^2/(V_t^2 + V_b^2) \approx 0.9$ was required to fit histograms (see Figure S10 in SI), as expected from the structure of the molecule and previous studies[12,14]. The "large" values of $V_t$ and $V_b$ (molecular orbital energy broadening amounts of 100 meV and 15 meV, respectively) confirm our expectation of strong molecule/electrode couplings, which we previously exploited to obtain a high-frequency molecular diode[14].

Extracted distributions of $t$ corresponding to best fits in Figure 5d are shown in Figure 6a. We have explored two $t$ distributions corresponding to eq 1a and eq 1b. In both cases, maxima are found at $t \approx 35$ meV, which is in the expected range from our DFT calculations. However, both deviate quantitatively from the theoretical distribution prediction for $t$ based on thermal molecular motions (SD($t$) $\approx$ <$t$> in eq 1a).[25] Using eq 1a, we find SD($t$) $\approx$ 140 meV, suggesting that the structural fluctuations are larger than those generated from solely thermal motions (phonons). Structural fluctuations are explicitly considered with parameter $\delta d$ in eq 1b. The extracted SD($\delta d$) = 0.8Å is reasonable given that a more packed configuration for these monolayers is possible.[12] Based on these results, we suggest that Van der Waals interactions between alkyl chains, which compete with the π-π interactions in the molecular organization of such monolayers,[12] could play a role in the distribution of $t$.

We stress that the number of molecules $NxN$, considered for current histograms generation, affects the quantitative extraction of $t$. An approximately 150 molecules are used in the experiment. A large enough $N$ was required in the model to avoid overestimating the extracted value of $t$ (Figure S11 in SI). At $N$=9, the extracted $t$ depends to a lesser extent on the molecule/electrode coupling parameters, which reduces the error on the estimated $t$ (Figure S11 in SI). We suggest that $t \approx 35\pm20$ meV is extracted from the present model based on both $t$ distributions and the possible error on $V_t$ and $V_b$.

From this quantitative analysis on *t*, we can discuss the results in the general contexts of charge transport in organic semiconductors[32,33,68] and chemical characterization tools.

As high-mobility organic semiconductors are often composed of a π-conjugated backbone substituted by one or more alkyl side chains,[32] as in the present study, a *t* distribution following eq 1b may be considered in charge transport models. Semiclassical theories of charge transport in organic semiconductors show that the electron transfer (hopping) rates along the π-conjugated molecular planes scale as $t^2$. Figure 6b represents such probability distributions for $t^2$ corresponding to the two *t* distributions shown in Figure 6a (related to eq 1a and 1b). Distributions have similar shapes in both cases, but the tail is narrower for the Gaussian distribution of *t*. In both cases, the broadened distribution of $t^2$ would open new hopping pathways.

The exploration of π-π intermolecular interaction energies from both CV and current histograms using the same samples composed of a large array of Fc-thiolated gold nanocrystals enables a direct comparison of both techniques as chemical characterization tools. Parameters *φ* and *t* are different in nature, but both are related to the molecular organization. As for *t* with eq 1b, *φ* can be related to *δd* from a simple electrostatic model (Figure 6c, inset):

$$\varphi = q.[1-(1+(r_a/(d+\delta d))^2)^{-0.5}]/[4\pi\varepsilon_0\varepsilon_r(d+\delta d)] \quad (2)$$

$r_a$ is the counter-ion pairing distance, *q* is the elementary charge, $\varepsilon_0$ and $\varepsilon_r$ are the dielectric permittivity of vacuum and the relative permittivity of water, respectively. *φ*=4.5 meV, for *d*=7 Å and *δd=0* Å, corresponds to an Fc-ClO$_4^-$ ion pairing distance of 4.9 Å (5.5Å is expected from molecular dynamics simulations[69]). With eq 2, a Gaussian distribution for *δd* implies a non-Gaussian distribution for *φ* (Figure 6c). Combining eq 2 and the extended Laviron model (see SI Methods), we see that such a distribution should induce a broadening of the CV peak, but only when $d+\delta d$ approaches the ion-pairing distance (Figure 6d). Therefore, CV would not be sufficiently sensitive to assess information on the small molecular organization fluctuations expected here (e.g. SD($\delta d$) = 0.8Å from parameter $t$ analysis). This feature illustrates the potential of using $t$ as an ultra-sensitive chemical characterization parameter.

In summary, we have investigated the possibility of assessing the $\pi$-$\pi$ electronic couplings from charge transport measurements in a connected device, using a statistical analysis of current from a large array of Fc-thiolated gold nanocrystals. The results have been quantitatively compared to DFT calculations. Extracted parameters, including a molecule/electrode coupling asymmetry $\alpha$ of 0.9 and $t$ of 35 meV, were in the range of expectations. However, the distribution of $t$ was broader than expected from the solely thermal fluctuations. This observation is attributed to structural fluctuations and to a variation of the intermolecular distance of 0.8 Å in the model. The results confirm the need for charge transport model to consider small structural fluctuations, even on the order of 1Å; however, CV does not have sufficient sensitivity to reveal such small fluctuations. This limitation may be overcome by measuring extremely small CV currents (on the single-dot level), and performing statistical analyses on $\varphi$ (as predicted in Figure 6c). The origin of these structural fluctuations remains unclear, but could be related to the competitive $\pi$-$\pi$ and $\sigma$-$\sigma$ interactions due to the presence of alkyl chains. Overall, the present study provides insights into understanding $\pi$-$\pi$ intermolecular interactions in organic and (bio-) molecular devices. The findings confirm that Landauer-type coherent-scattering models, which are usually dedicated to low-temperature mesoscopic physics, are relevant at room temperature for molecular electronics, even in the presence of cooperative effects. Statistical current analysis could be applied to various systems, because current histograms represent a common approach in molecular junctions. The study of π-π electronic couplings is a unique opportunity to link quantum chemistry, mesoscopic physics, organic electronics, and electrochemistry, indicating the importance of each subfield in the development of organic electronics.

**Methods**

Additional Methodological information related to STM, gold nanodot fabrication, monolayer self-assembly, experimental conditions for CV and related fits, image treatment DFT calculations and theoretical histogram generation is available in the SI Methods.

**UHV STM**

The high resolution image was performed at room temperature with a substrate biased at 2V and at a constant current of 1 pA.

**Areas of top and side facets of the nanocrystals**

To estimate the number of molecules per peak, the area was considered from the following formula, based on Figure 2c:

$$\pi/4*(D-2*h/tan\xi)^2 + \pi*(D-h/tan\xi)*h/sin\xi \qquad (3)$$

where the first term corresponds to the area on the top and the second term to the area on the sides of the nanocrystal. Reasonable fits are obtained with $h$=2.7 nm and $\xi$=30°, as expected from the nanocrystal structure.

**C-AFM**

We measured current voltage by using C-AFM (Dimension 3100, Veeco) with a PtIr-coated tip on molecules[66] in $N_2$ atmosphere. Each count in the statistical analysis corresponds to a single and independent gold nanodot. The tip curvature radius is about 40 nm (estimated by SEM), and the force constant is in the range of 0.17—0.2 N/m. C-AFM measurements were taken at loading forces of 15 and 30 nN for the smallest and largest dots, respectively to keep a similar force per surface unit. As shown in ref.14, a weak effect of the force is observed for these molecules in the range of 10—30 nN. In scanning mode, the bias is fixed and the tip sweep frequency is set at 0.5 Hz. With our experimental setup being limited to 512 pixels/image, the parameters lead to a typical number of 3000 counts for a 6×6 μm C-AFM image. In the presence of $\pi$-$\pi$ electronic couplings, current histogram peaks are well-fitted with an asymmetric double sigmoidal function $f(x)$ given by

$$f(x) = y_0 + A \frac{1}{1+e^{-(x-x_c+w_1/2)/w_2}} \left(1 - \frac{1}{1+e^{-(x-x_c-w_1/2)/w_3}}\right) \quad (4)$$

where the values of the various parameters are presented in SI Table 2.

**Landauer Imry Buttiker Formalism and histograms fits**

The model has been adapted from ref.21 to account for a large number of molecules and asymmetrical contacts (detail in SI Methods). It provides a good description of the fundamental aspects of electron transport via the computation of the energy-dependent transmission through the device. The formalism described in ref.21 focused on the zero-bias conductance (and at low temperature), a result which can be extended to the evaluation of the current at low bias, provided that one integrates the transmission over a range of energy given by the external potential. Here we have used this model with conditions of relatively high bias and at room temperature. The assumption that the Landauer approach remains applicable under such conditions is often made in the field of molecular electronics with a single-level model.[31] We believe that these assumptions are further justified due to the strong coupling of the Fc molecules, as evidenced by the level broadening estimated in the range of 100 meV. The generation of current histograms (instead of conductance histograms in ref 21) required an additional assumption (midpoint rule) to efficiency compute the $10^6$ realizations (see SI Methods). The validity of this approximation has been confirmed for the present study (Figure S14 in SI).

The process of fitting the line shape of the experimental histograms relies on a relatively large number of variables, which can be defined by a step by step procedure. We considered a site energy $\varepsilon = 0.2$ eV versus Fermi level at $V_{bias}=0$ V, given the CV results and related energy band diagram proposed in ref. 14. An upper limit of $\delta\varepsilon<45$ meV was considered based on CV analysis (Figure S3d). First, we optimized the parameters for 3×3 molecules due to computational time. Because $\delta\varepsilon$ does not significantly affect the extracted value of $t$ (Figure S10), we considered $\delta\varepsilon=40$ meV (similar to the value considered in ref.21. $V_t$ and $V_b$ were adjusted to get a good current level and to reproduce the histogram shape. An optimal asymmetry factor of $\alpha=0.9$ was considered (Figure S10), in agreement with refs. 12 and 14. $\delta V$ was tuned to fit the histogram line shape when $t = 0$ (mixed monolayer). $\bar{\delta}t$ was tuned with $t$ while fitting asymmetric histograms line shapes. Current histograms were generated based on $10^6$ realizations. When the number of molecules in the matrix is large (e.g. 9x9 molecules), histogram fitting takes several days. Efficient hardwares (i.e. Ising machine) is being developed to solve such problems efficiently.[70,71]

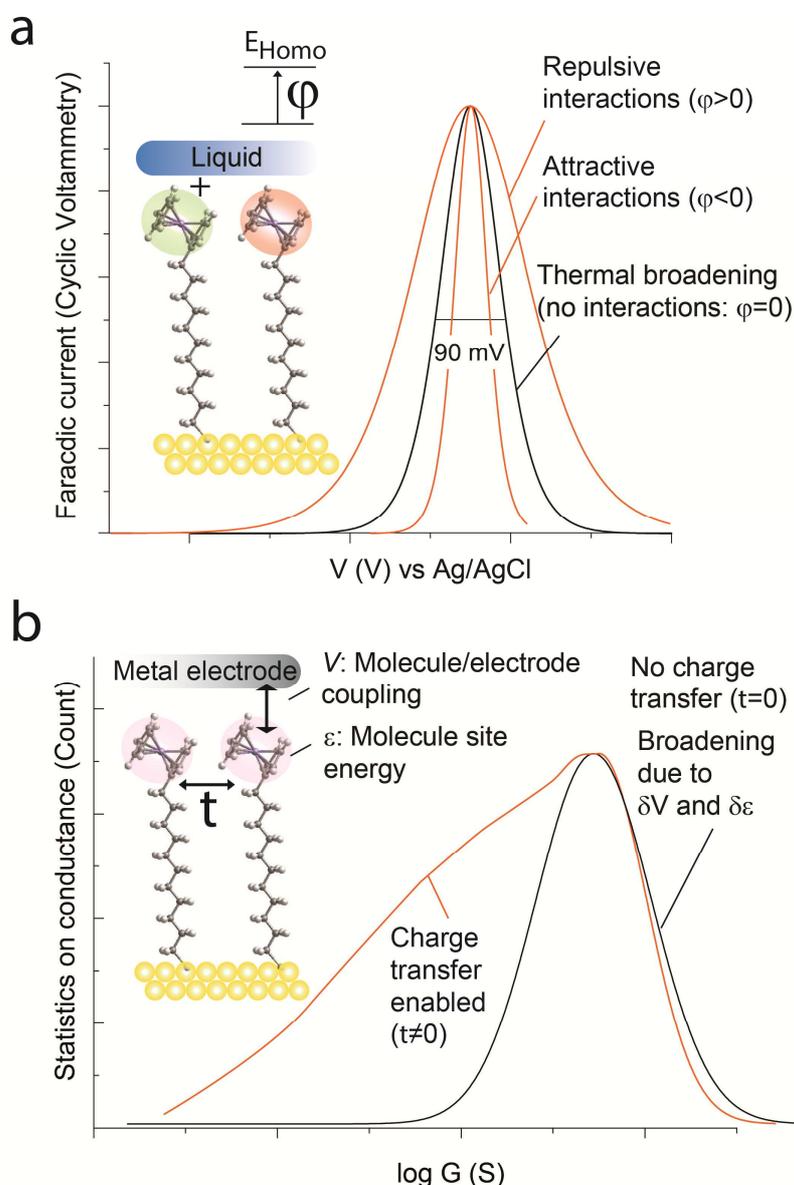

**Figure 1 Signatures of cooperative effects with the introduction of parameters $\varphi$ and $t$.** (**a**) Schematic representation of CV results in the absence (black curve) and presence (orange curve) of Coulomb interactions between Fc molecules according to the Laviron model[27] based on the Frumkin isotherm. Inset: Schematic representation of the microscopic process. When Fc is oxidized (green cloud), it shifts the energy level of the neighboring molecule by $\varphi$. (**b**) Schematic representation of a theoretically predicted[21] conductance histogram in the absence (black curve) and presence (orange curve) of coupling between two molecules (tight binding model). Inset: Schematic representation of intermolecular coupling ($t$) related to charge transfer between adjacent molecules.

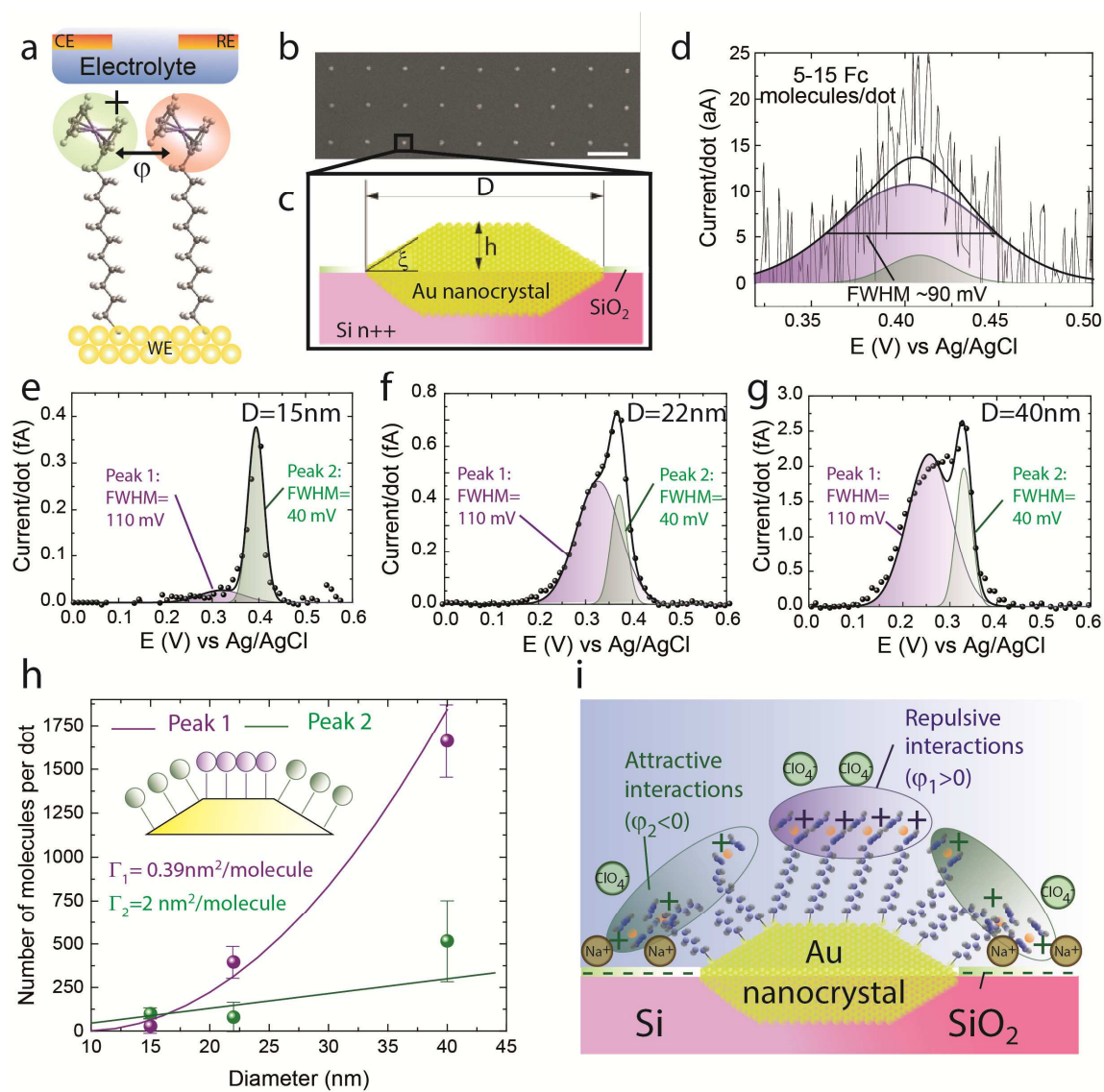

**Figure 2 Quantitative analysis of intermolecular interaction energy $\varphi$ from CV**
(**a**) Schematic representation of intermolecular interaction in an electrochemical setup. Electrolyte is $NaClO_4$ (0.1 M). $\varphi$, Coulomb repulsion between adjacent molecules (see SI Methods); CE counter electrode; RE Ag/AgCl reference electrode; WE working electrode (gold nanocrystals). (**b**) SEM image of gold nanocrystal array on highly doped silicon. Scale bar = 200 nm. (**c**) Schematic representation of a nanocrystal cross-section based on ref.49. Parameters of interest are the nanocrystal diameter $D$, height $h \approx 3$ nm and angle $\xi \lesssim 30°$. (**d**) Square wave voltammogram (SwV) for (1:10) $FcC_{11}SH$ SAM diluted with $C_{12}$ molecules on an array of 15 nm diameter dots. Integration of the main peak area corresponds, after normalization (see SI Methods), to 5 to 15 $FcC_{11}SH$ molecules per dot. The curve is fitted with 2 peaks. The main peak has a FWHM~90 mV ($\varphi \approx 0$). (**e-g**) CV results (anodic peak) for arrays of $FcC_{11}SH$-coated gold nanoelectrodes of different diameters (as indicated). FWHM correspond to sweep rate of 1V/s (oxidation peak). Fitting parameters are indicated in Table S2.

**(h)** Graphs showing number of molecules per dot (obtained from (e-g)) averaged with data from reduction peak (Figure S6). Error bars are based on dispersion between CV (oxidation/reduction peak and various speeds). Data are fitted with eq 3 (truncated cone approximation). Inset: Schematic representation of molecular organization. Peak 1 (purple) and peak 2 (green) correspond to molecules on top and sides, respectively. (**i**) Schematic representation of Fc-thiolated gold nanocrystals in NaClO$_4$ electrolyte when all Fc are oxidized.

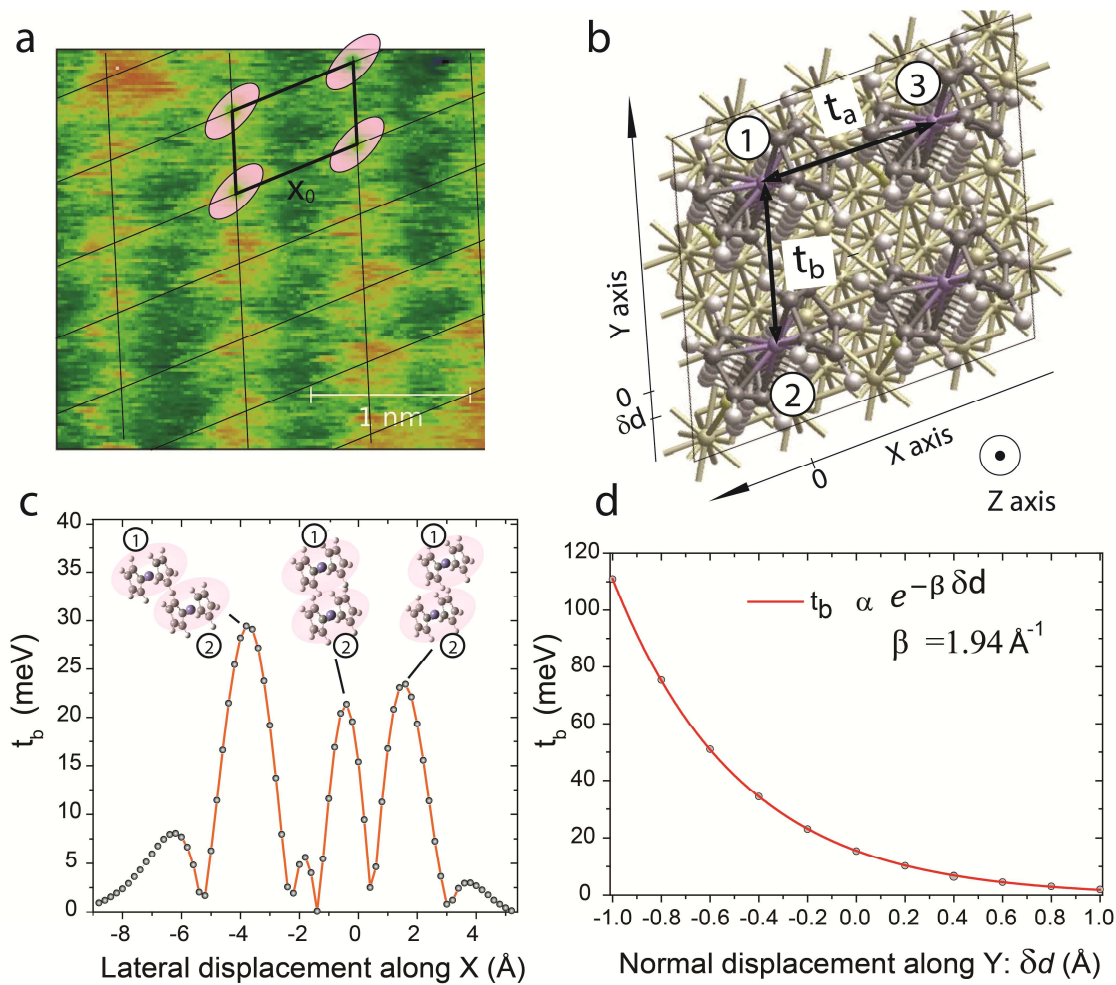

**Figure 3 Estimation of cooperative effects from supramolecular organization and DFT calculations** (a) UHV-STM image of a SAM of $FcC_{11}SH$ molecules grafted on gold. Molecular structure is resolved and used as reference for full DFT calculations. Periodic black lines, with cell delimited by pink clouds, indicate positions of Fc molecules. (b) Cell composed of four $FcC_{11}SH$ molecules based on (a,b). A number is attributed to each molecule due structural anisotropy. $(\varphi_a, t_a)$ and $(\varphi_b, t_b)$ refer to interactions between molecules 1 and 3 and molecules 1 and 2, respectively. X and Y axes are aligned along molecules 3 and 1 and molecules 2 and 1, respectively. (c) Full DFT calculation of parameter $t_b$ between molecules 1 and 2. Position of molecule 2 is translated along the X axis to mimic disorder. Inset: Molecular configuration at each maximum for $t_b$. (d) Evolution of $t_b$ as a function of the variation of intermolecular separation $\delta d$ modulated from the initial geometry (normal displacement = 0 Å). Decay ratios $\beta_b$ is indicated.

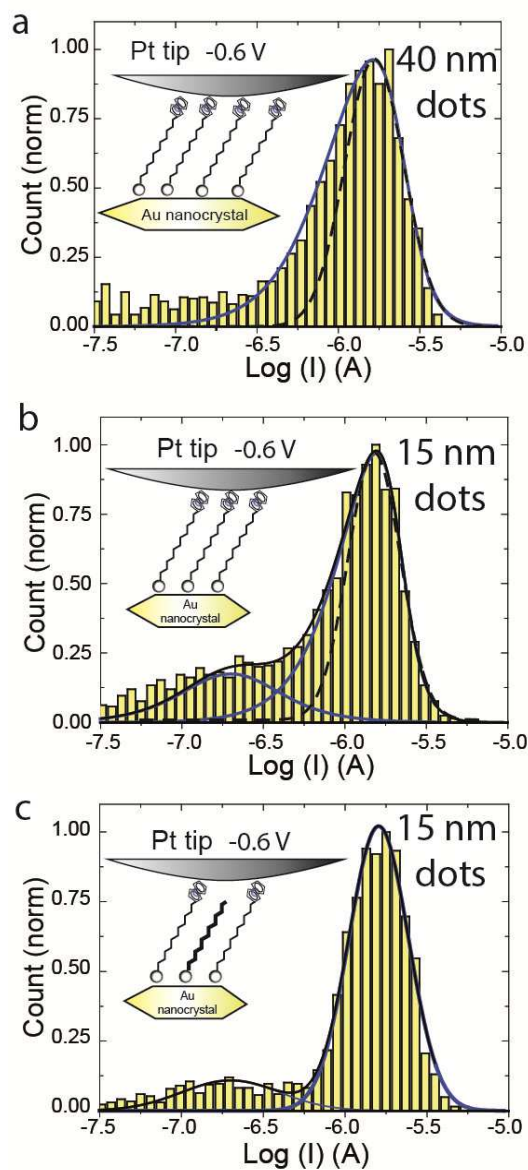

**Figure 4 Current histograms used to evaluate π-π intermolecular interaction energy (~ 3000 counts per histogram)** Current histogram obtained at a tip voltage of -0.6 V for **(a)** 40 and **(b)** 15-nm diameter dots (with 5-nm diameter on top). Plain curve is the fit with asymmetric double sigmoidal function (eq 4). Dashed curve is the log-normal fit. Inset: Schematic view of the setup. **(c)** Same as (b) but with a (1:1) $FcC_{11}SH:C_{12}SH$-diluted SAM. Fitting parameters are shown in Table S2.

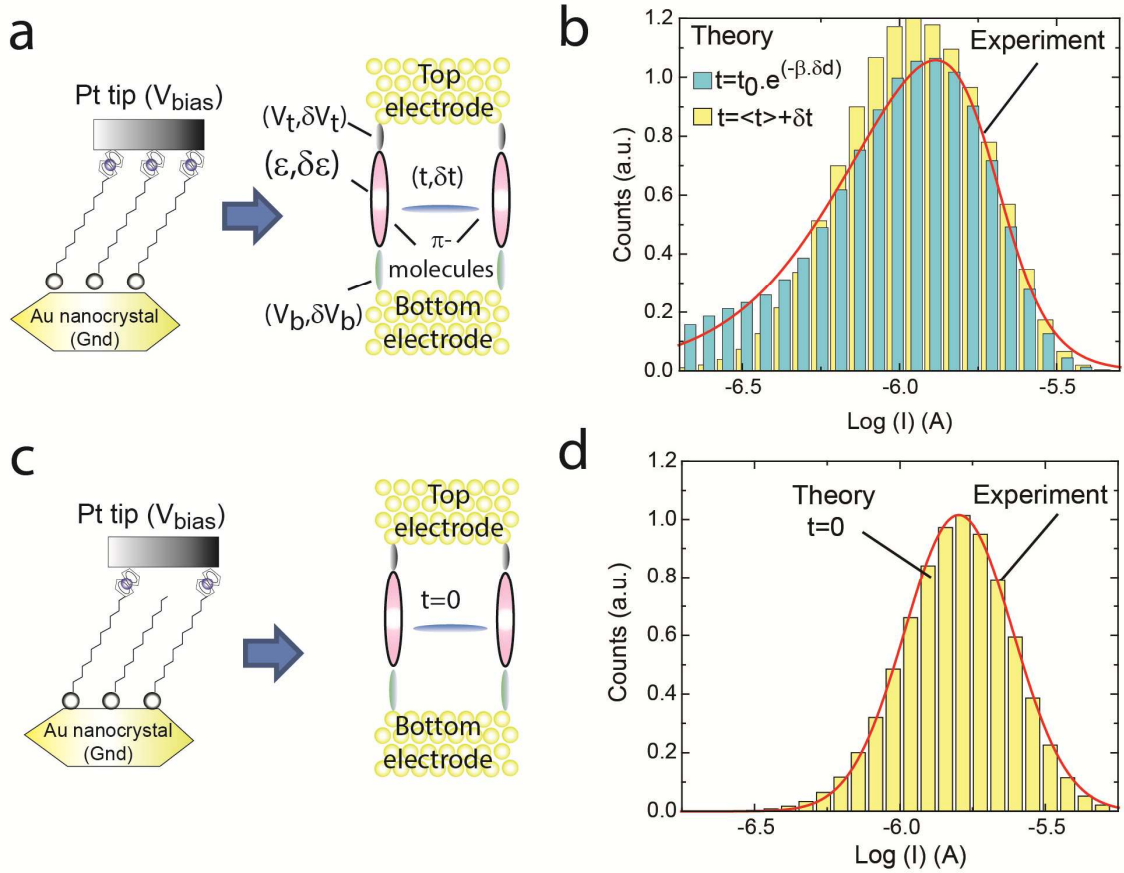

**Figure 5  Histograms fits with Landauer Buttiker Imry formalism.**
(**a**) Schematic representation of the model. Each molecule (quantum dot) is coupled to other molecules with coupling term *t* and coupled to top/bottom electrodes with coupling energies $V_t$ and $V_b$, respectively. ε, molecule orbital energy. SDs of these parameters are used to generate histograms. Related experimental setup shown for clarity. (**b**) Experimental ($V_{tip}$ = -0.6 V) and simulated histograms ($V_t$ = 0.401eV, $V_b$ = 0.144eV, $\delta V$ = 22meV, $\varepsilon$ = 0.2eV, $t$ = 0.04 eV, $\delta t$ = 140 meV, $t_0$ = 0.34eV, $\beta$ = 1.96/Å SD($\delta d$)=0.8Å) considering 81 molecules. (**c**) Schematic representation of model with fewer molecular interactions. Parameters are same as in (**a**) except $t$ = 0. Related experimental setup (diluted monolayer) is shown for clarity. (**d**) Experimental ($V_{tip}$ = -0.6 V) and simulated histograms ($t$ = 0).

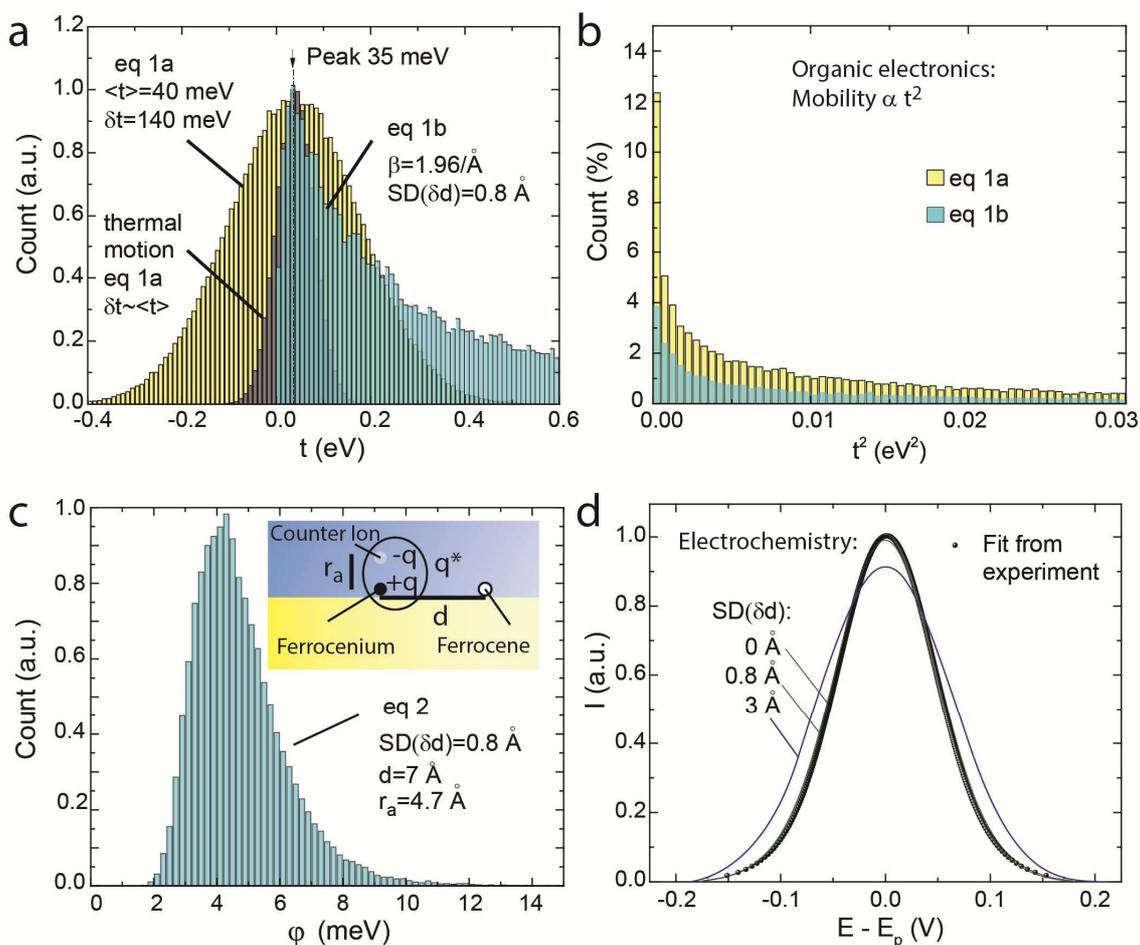

**Figure 6 Extracted distributions of *t* and implications for organic electronic and electrochemical field** (**a**) Distribution of *t* obtained from best fits in Figure 5d with eqs 1a and 1b. Expected (Gaussian) distribution from solely thermal motions shown in grey. Each electronic level has an associated *t* that can be positive or negative, so the sign is of little importance. (**b**) $t^2$ distribution obtained from (a). (**c**) Estimated distribution of $\varphi$ from eq 2 given an intermolecular distance fluctuation of $\delta d$=0.8Å. Inset: Schematic representation of the electrostatic model (equation 2). (**d**) Experimental and theoretical CV results (coupled eq 2 and eq S1) with $\delta d$=0, 0.8 and 3 Å. Energy level of Fc vs Ag/AgCl ($E_p$) is used to center the CV peak at 0.

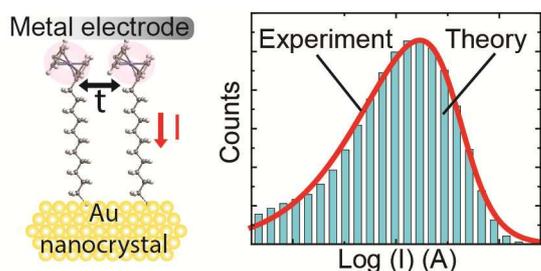

**ToC Image** The π-π electronic coupling energy parameter *t* is assessed from a statistical analysis of current histograms.

**Associated content:**

**Supporting information**

The Supporting Information is available free of charge on the ACS Publications website at DOI:10.1021/acs.nano-lett.xxx

**Corresponding Author:**

e-mail : nicolas.clement@lab.ntt.co.jp

**Acknowledgments**

The authors thank C.A. Nijhuis from NTU Singapore for discussions, C. Wahl from CPT for beginning simulations on current histograms, D. Guerin and A. Vlandas from IEMN for discussions related to electrochemical measurements, and T. Hayashi, K. Chida and T.Goto from NTT Basic Research Labs for fruitful discussions. J.T. thanks PhD funding from Marie Curie ITN grants and the EU-FP7 Nanomicrowave project and J.C thanks the iSwitch (GA No. 642196) project. We acknowledge support from Renatech (the French national nanofabrication network) and Equipex Excelsior. Work in Mons was supported by the Interuniversity Attraction Pole program of the Belgian Federal Science Policy Office (PAI 7/05) and by the Belgian National Fund for Scientific Research (FNRS). JC is a research director of FNRS.28

---

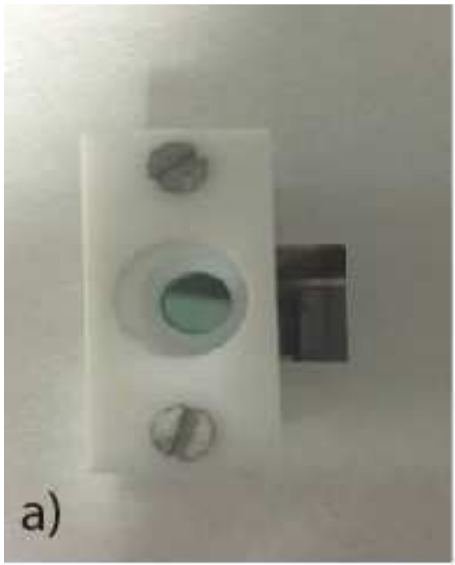 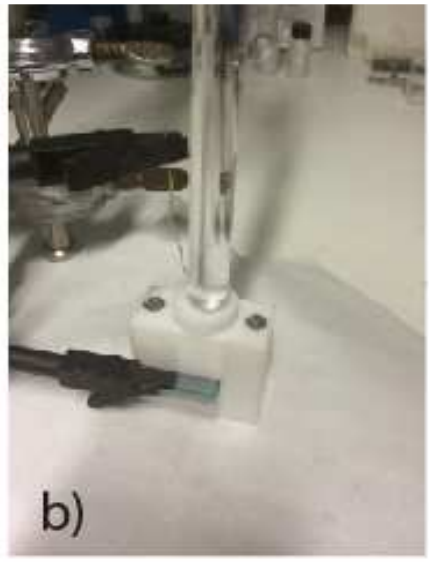

**Supplementary Figure 1 a.** Electrochemical cell with a sample. **b.** Setup during the operation.

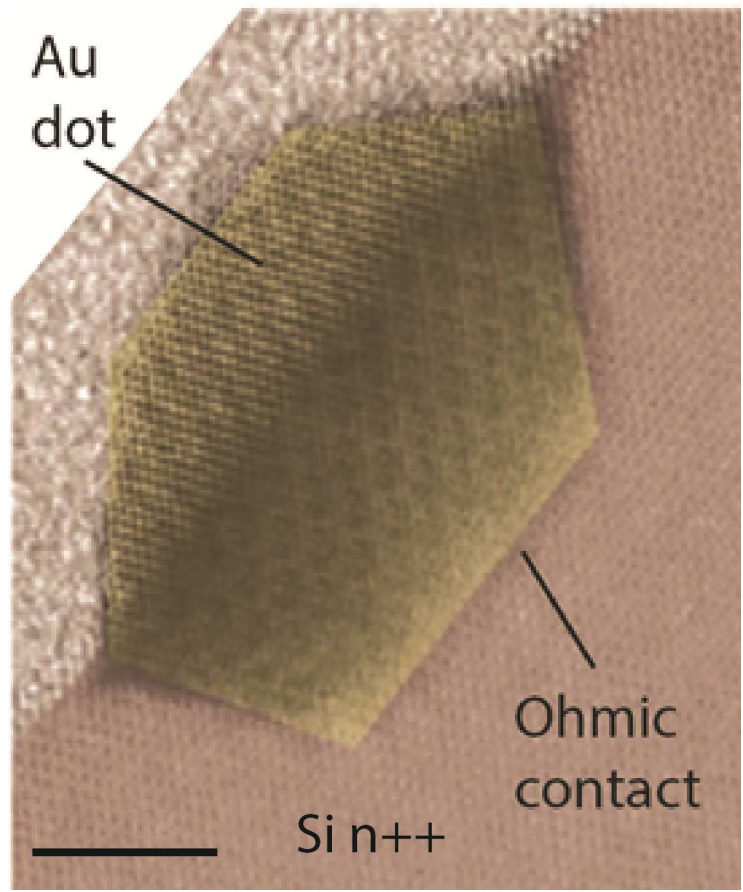

**Supplementary Figure 2.** Gold nanodot electrode fabrication Coloured TEM images of an Au nanodot (adapted from Ref.[1]). Scale bar is 5 nm.

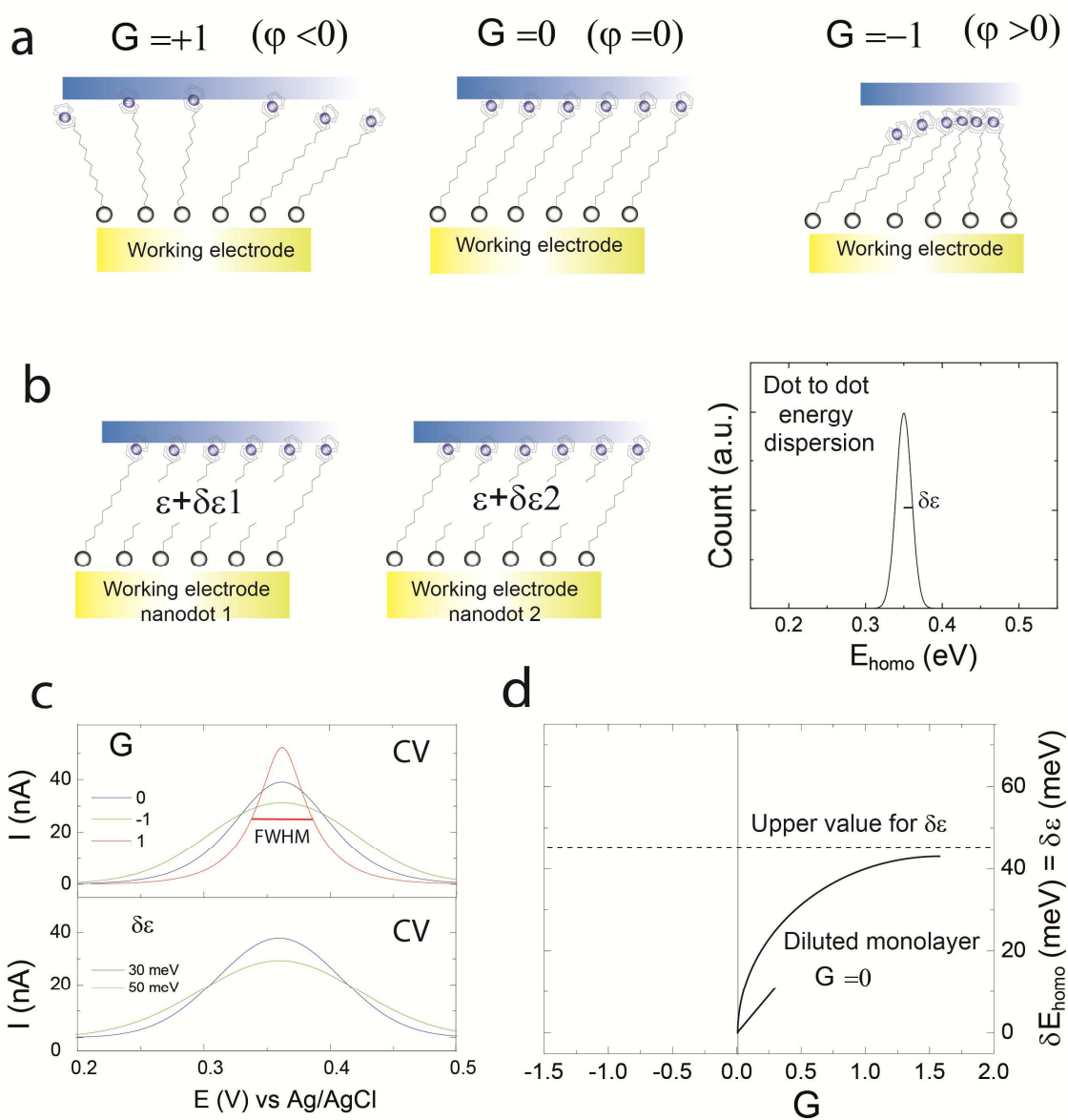

**Supplementary Figure 3 a** Typical "oversimplified" schematic representation for Coulomb interactions between redox moieties and the related value of *G*. **b** Schematic representation of the dot to dot dispersion in the energy position for the HOMO level. We write $\delta\varepsilon$ instead of $\delta E_{HOMO}$ for simplicity and for consistency with conductance statistics analysis. **c** Graphs showing the impact of *G*. and $\delta\varepsilon$ parameters on cyclic voltammograms line shape. FWHM represents the full width at half maximum. **d** Graph showing the correlation between *G* and $\delta\varepsilon$ parameters as there is always a couple (*G*, $\delta\varepsilon$) of solutions to fit datas. We see that *G*~0 implies an upper value for $\delta\varepsilon$=45 meV.

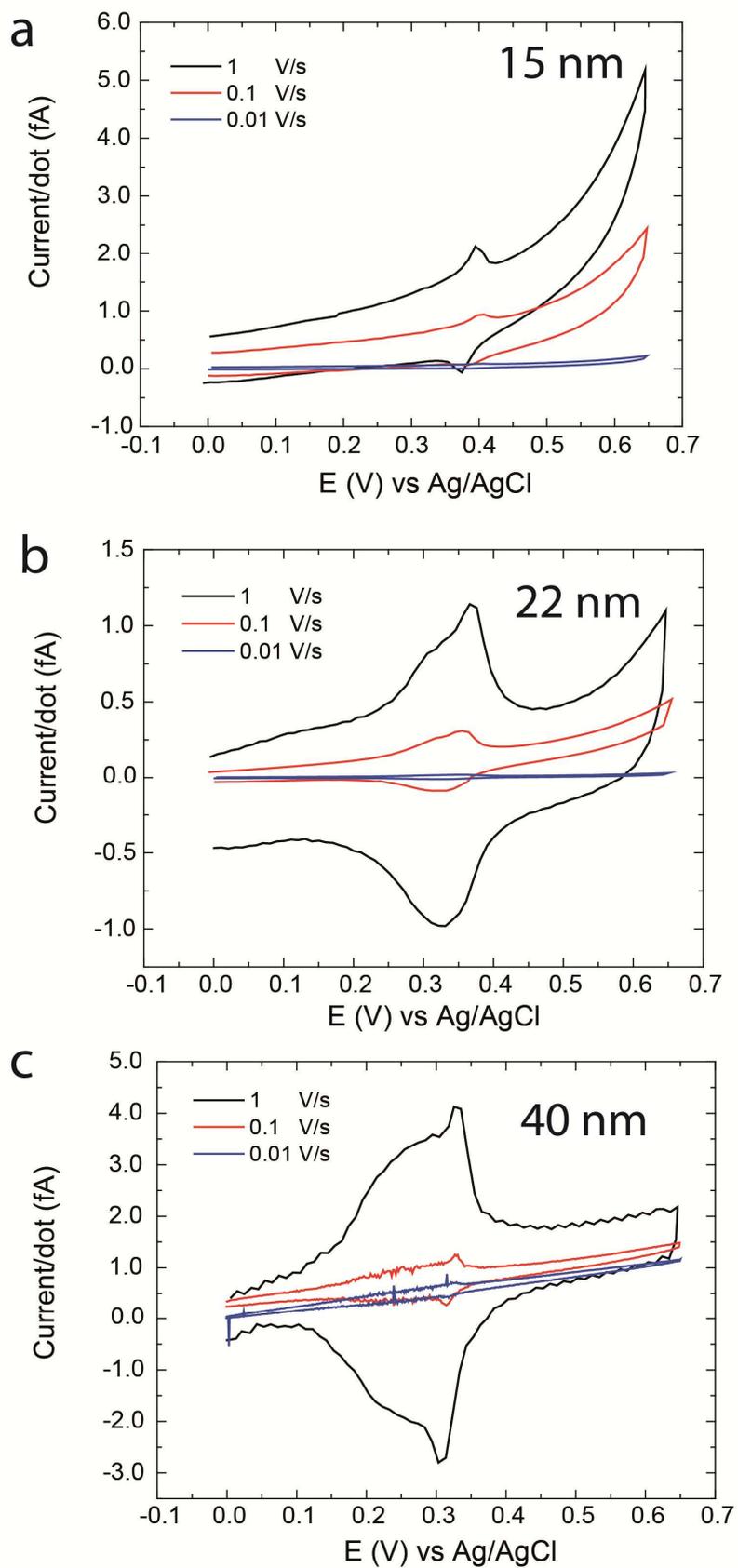

**Supplementary Figure 4 Raw CV measurements at 0.01, 0.1 and 1 V/s and different dots diameters a** for 15 nm diameter dots. **b** for 22 nm diameter dots. **c** for 40 nm diameter dots.

**Supplementary Figure 5 CV after baseline removal at 1 V/s for different dots diameters (oxidation and reduction peak) a** for 15 nm diameter dots. **b** for 22 nm diameter dots. **c** for 40 nm diameter dots.

**Supplementary Figure 6 CV after baseline removal for 22nm-diameter dots at different sweep rates (oxidation and reduction peak) a** at 1 V/s **b** at 0.1 V/s **c** at 0.01 V/s. **d** Summary of the number of molecules per dot obtained for each peak and the three different sweep rates.

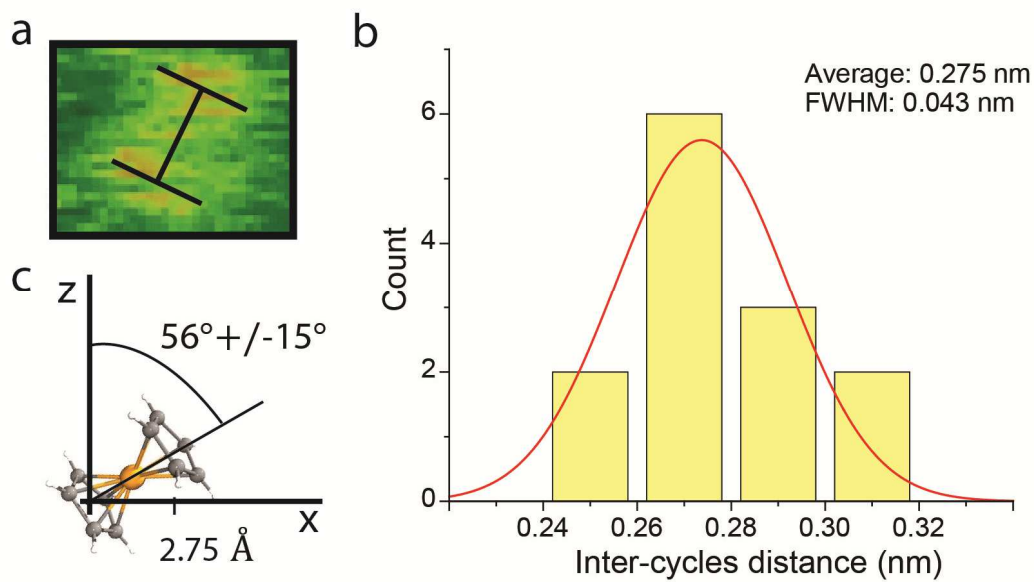

**Supplementary Figure 7 Estimation of the tilt angle from the STM image a** Example of intercycle extraction from the STM image shown in Fig.1d. In this example, the intercycle distance is 0.27 nm. **b** Histogram of the estimated Inter-cycles distance from Fig.1d. c Schematic representation of the molecule tilt angle formed by the Fc and the normal axis (z), based on b. For this estimation, we considered an iron-cycle distance of 1.664Å based on [2,3].

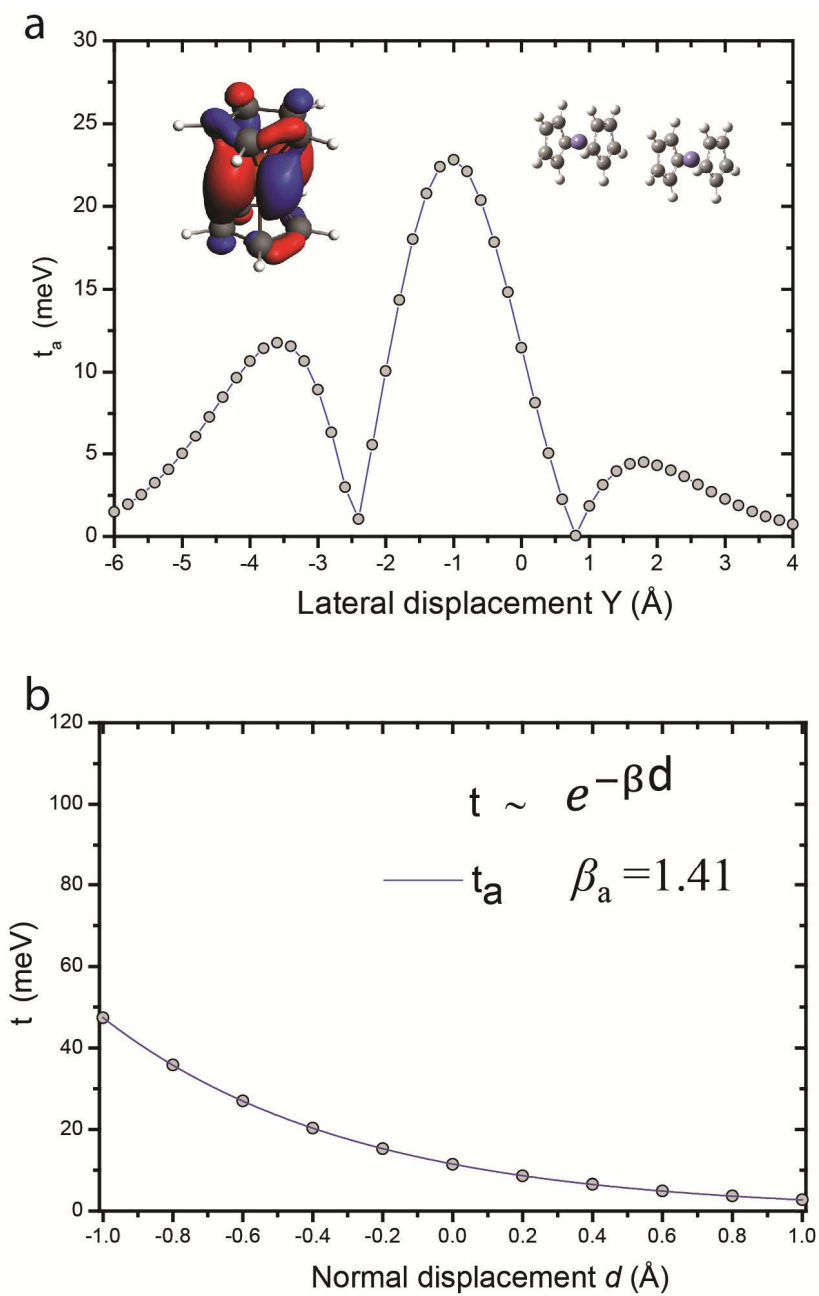

**Supplementary Figure 8 Additional DFT calculations a** DFT calculation of parameter $t_a$ between molecules 1 and 3. The position of the molecule 3 is translated along the Y axis to mimic disorder. At the maximum for $|t_a|$, the related molecular configuration is shown in inset. The shape of the HOMO molecular orbital for the Fc moiety is also shown in inset. **b** Evolution of the transfer integral as a function of the intermolecular separation $d$ modulated from the initial geometry (normal displacement = zero Å). The decay ratios $β_a$ related to $t_a$ is indicated in the figure. This values is close to the tunnel decay ratio considered in molecular electronics.

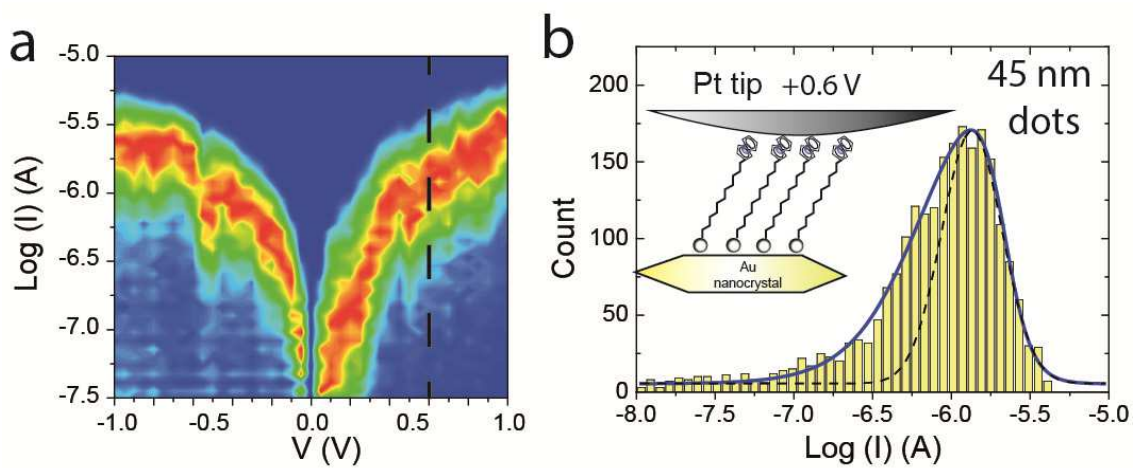

**Supplementary Figure 9 a** 2D current experimental histogram obtained with $FcC_{11}SH$ molecules on 45 nm dots. The dashed curve indicates +0.6V. **b** Current histogram obtained at tip voltage of +0.6 V for 45 nm dots. Plain curve is the asymmetric log-normal fit and dashed curve is the related symmetric log-normal curve. Inset: Schematic view of the setup.

**Supplementary Figure 10 Optimization of the various parameters for 3x3 molecules and a Gaussian distribution of t. a** Graph showing the impact of $\delta\varepsilon$ parameter on the extracted value of $t$ from fits. $\delta\varepsilon$ has a weak impact on $t$. **b** Graph showing the impact of $V_t$ and $V_b$ parameters on the extracted value of $t$ from fits. The error to fits is shown in **c,d**. Values corresponding to minimum fitting error are indicated by dashed lines (zone in between both lines) and by a blue area. The minimum in **c** is used to choose the value of $V_t$. There is no clear minimum for $V_b$, but value giving good results for all the study, including the current level, is 0.144. It corresponds to a parameter $\alpha=0.9$. **e,f** Effect of coupling asymmetry in histograms line shape. Only a large value of α provides reasonable fits.

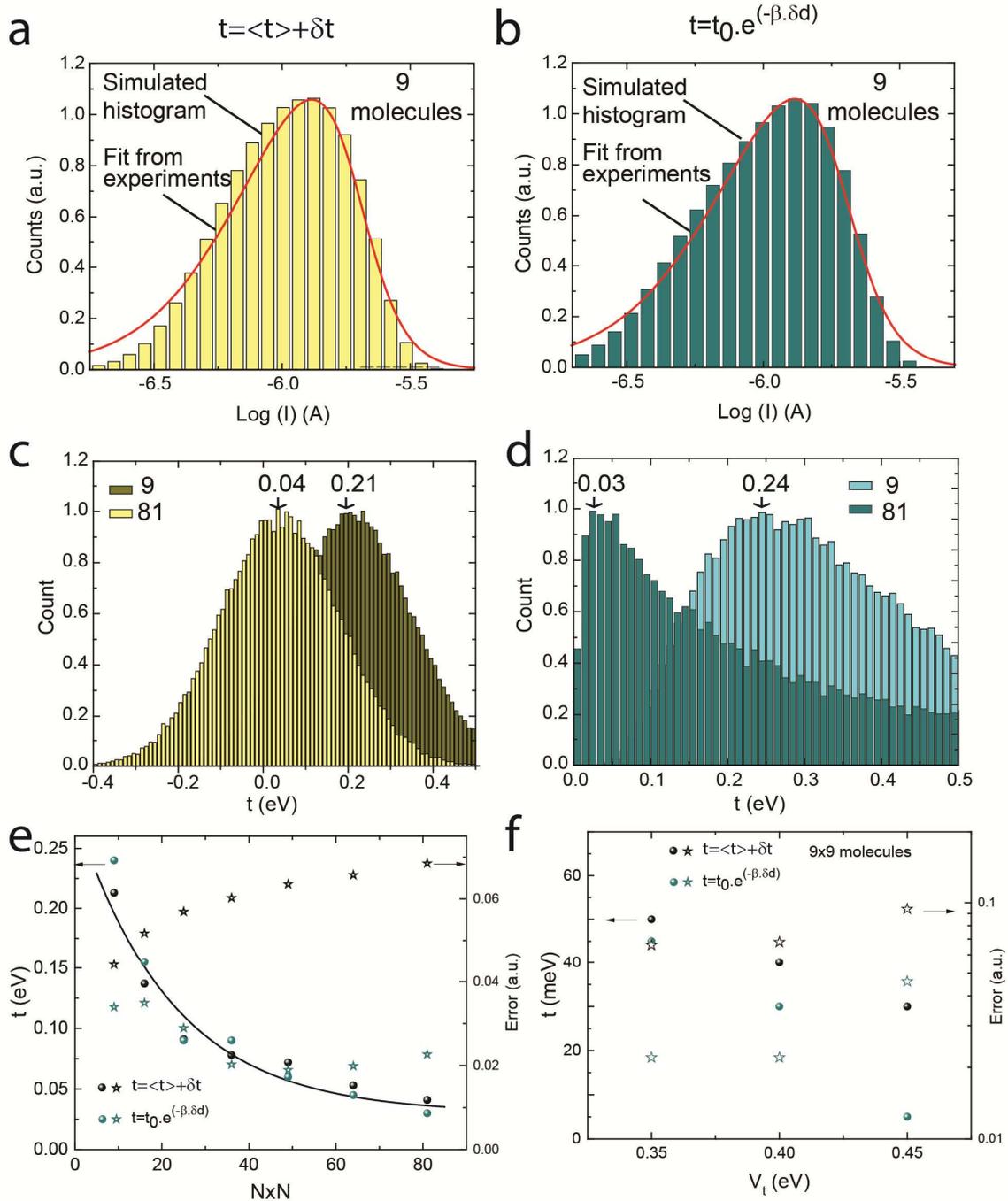

**Supplementary Figure 11 Impact of the number of molecules considered in the model in the extracted distributions of t. a** Histogram generated from the model with 9 molecules and with consideration of a Gaussian distribution of *t*. It corresponds to <*t*>=0.21 eV and *δt*=0.14 eV. The red line indicates the experimental reference data. **b** Histogram generated from the model with 9 molecules and with consideration of a Gaussian distribution of *d*. It corresponds to $t_0$=0.34 eV and *δd*=0.29 Å. The red line indicates the experimental reference data. **c** Distribution of *t* extracted from the best fits with a Gaussian distribution of *t*, and for 9 and 81 molecules in the model. **d** Distribution of *t* extracted from the best fits with a Gaussian distribution of *d*, and for 9 and 81 molecules in the model. **e** Graph showing the impact of the number of molecules considered in the model on the extracted value of *t* from the 2 distributions of *t*

investigated in a-d. In both cases, *t* depends exponentially on $N^2$ (with a decay factor of 0.045). The saturation is observed near *N*=9 (81 molecules). The plateau is at about 35 meV in both cases. This result does not mean that t depends on *N*, but that the appropriate number of molecules has to be considered in the model to quantitatively extract *t*. The error to the fit is shown in the right axis. We see that the Gaussian distribution of t gives the largest error, in particular at large *N*. **f** Dependence of estimated *t* for different values of $V_t$ in the case of 9x9 molecules. This result indicate a weaker dependence of *t* on molecules/electrode coupling (here $V_t$) when compared to a small number of molecules (see supplementary Fig.10). Also, the minimum error to fits corresponds to Vt=0.4eV (the error to fits is doubled when increasing $V_t$ to 0.45eV). Therefore, we can reasonably say that the extracted *t* is 35 ± 20 meV.

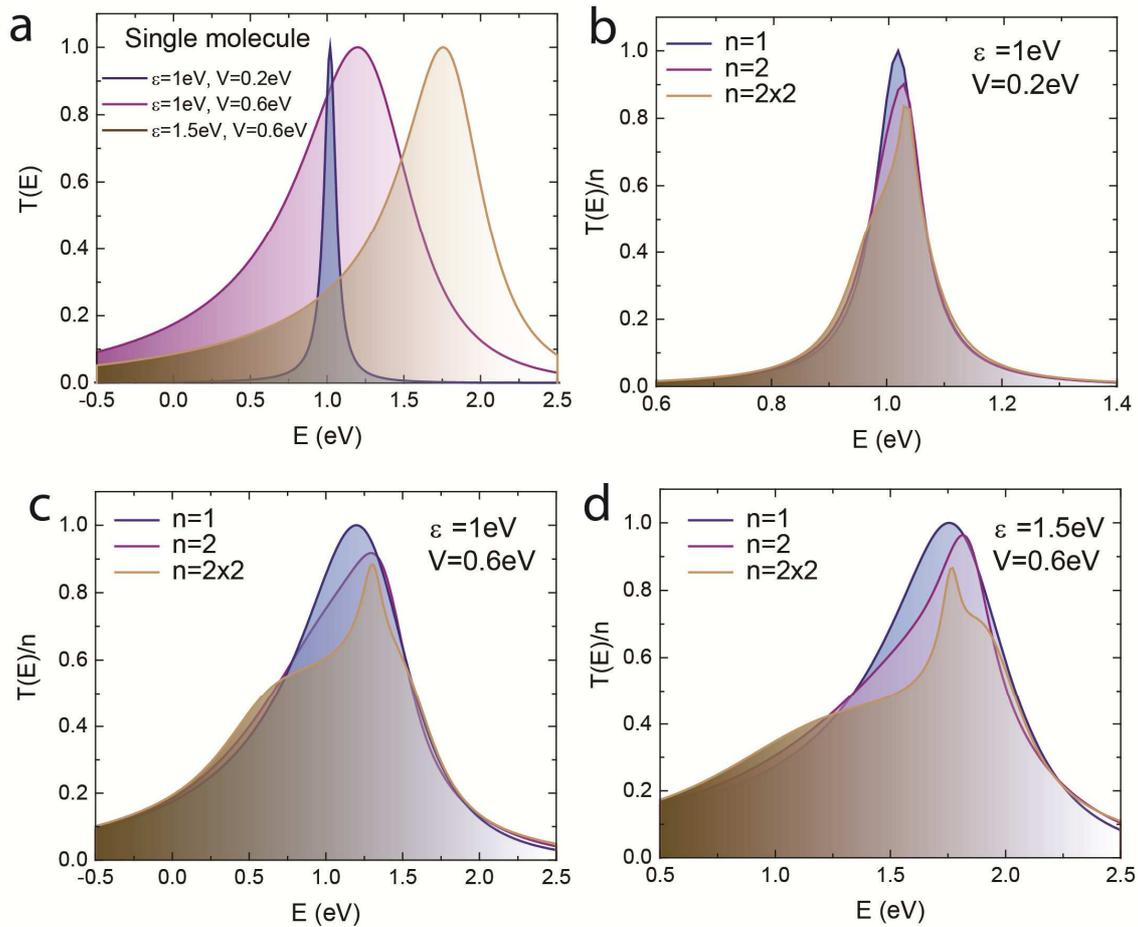

**Supplementary Figure 12 Description of cooperative effects on the Transmission coefficient for 1,2,2x2 molecules and *t*=0. a** Transmission as a function of the electrodes energy for a single molecule site at energy $\varepsilon = 1$, 1.5eV, and an electrode-molecule coupling $V = 0.2$, 0.6eV. For a single molecule, the conductance looks quite like a Lorentzian as a function of the energy of the electrodes. For a small electrode-molecule coupling, the transmission peak is narrow and centered around the molecule energy. When the electrode-molecule coupling is increased, the peak gets wider and it is also slightly shifted as the self-energy acquired because this coupling is larger. Finally, the transmission peak is drifted when the molecule energy is changed. **b-d** Average transmission per molecule as a function of the electrodes energy for n = 1, 2, 2x2 sites at energy $\varepsilon = 1$, 1.5eV, and an electrode-molecule coupling $V$ =0.2, 0.6eV. No direct charge transfer between molecules is allowed. When several molecules arranged in a wire couple the electrodes, a multiple path interference is observed, even when direct charge transfer between the molecules is forbidden. This effect does not depend much on the number of sites: the average transmissions per molecule T(E)/n are almost the same for n = 2 and n = 2x2.

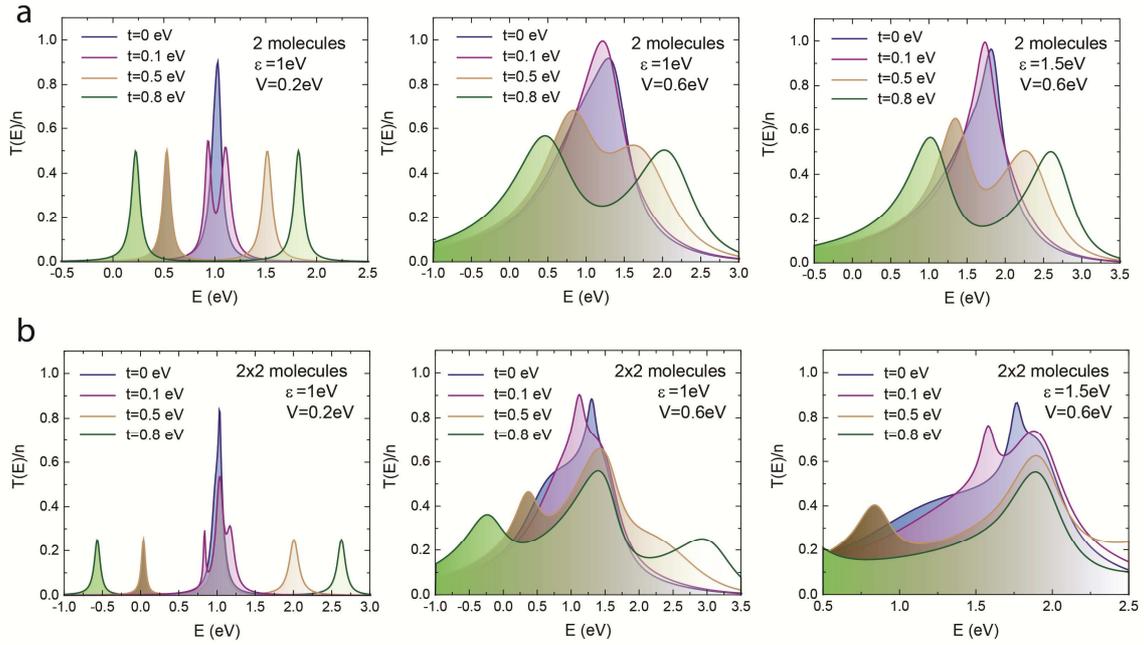

**Supplementary Figure 13 Description of cooperative effects on the Transmission coefficient for 2, 2x2 molecules and $t\neq 0$. a** Transmission as a function of the electrodes energy for n = 2 molecule sites arranged in a line at energy $\varepsilon = 1, 1.5$eV, and an electrode-molecule coupling $V = 0.2, 0.6$eV. When coupling between molecules is switched on, the transmission peak splits into 2 peaks. The peaks get more and more separated as transfer integral between molecules is increased. In fact, as many peaks as the number n of molecules appear. This can be seen when diagonalizing $H_{mol}$ (see supplementary Methods): the peaks observed in the transmission are centered around its eigenvalues. This diagonal basis for $H_{mol}$ is called the conduction channels basis. For example, at $n = 2$, the eigenvalues of the molecular Hamiltonian are $\varepsilon \pm t$, which explains why the peaks are drifted away when $t$ is increased. **b** Transmission as a function of the electrodes energy for n = 2x2 molecule sites arranged in a square at energy $\varepsilon = 1, 1.5$eV, and an electrode-molecule coupling $V = 0.2, 0.6$eV. When charge transfer between molecules is switched on, the transmission peak splits into 4 peaks. The peaks get more and more separated as the transfer integral is increased.

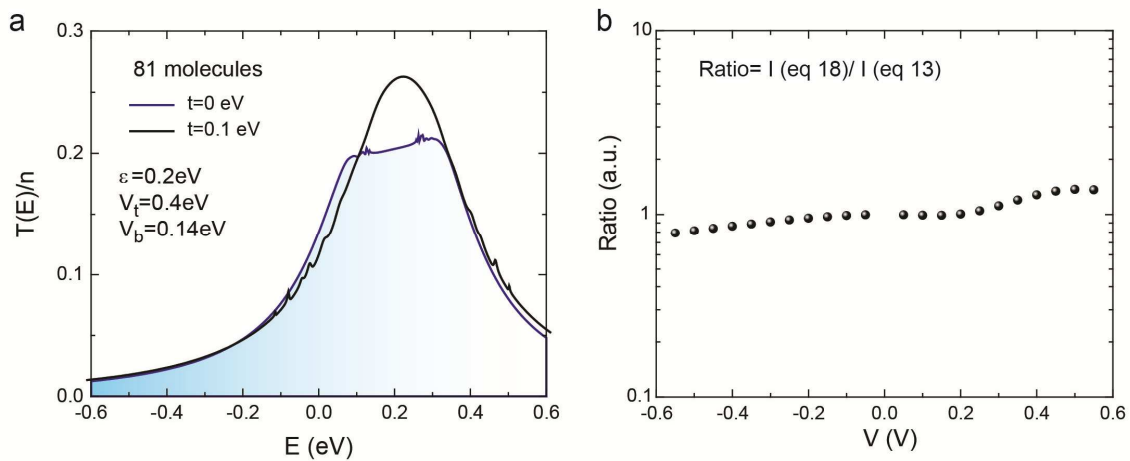

**Supplementary Figure 14 Transmission function and evaluation of the approximation error in current** (**a**) Transmission as a function of electrode energy for 9×9 molecules (with parameters used to fit histograms). (**b**) Ratio of the current estimated from eq 18 and 13 considering the T(E) shown in (a).

| Φ(nm) | $E_1$ (V) | FWHM$_1$ (V) | $G_1$ (-) | $E_2$ (V) | FWHM$_2$ (V) | $G_2$ (-) |
|---|---|---|---|---|---|---|
| 40 | 0.25 | 0.11 | -0.356 | 0.32 | 0.04 | 0.959 |
| 22 | 0.32 | 0.11 | -0.356 | 0.35 | 0.04 | 0.959 |
| 15 | 0.35 | 0.11 | -0.356 | 0.39 | 0.04 | 0.959 |

**Supplementary Table 1** Fitting parameters for CV. $Ep_1$, $Ep_2$ are the peak position (oxidation state, 1V/s) for peak 1 and peak 2, respectively. FWHM$_1$ and FWHM$_2$ correspond to the FWHM of peak 1 and peak 2, respectively, and $G_1$, $G_2$ Laviron interaction parameters (see supplementary Methods) used to obtain $\varphi_1$ and $\varphi_2$.

| Size (nm) | 40 nm | 40 nm | 15 nm | 15 nm |
|---|---|---|---|---|
| Voltage | 600 mV | -600 mV | -600 mV | -600 mV |
| Dilution Ratio | 1:0 | 1:0 | 1:0 | 1:1 |
| Y0 | 0 | 0 | 0 | 0 |
| Xc | -5.92 | -5.80 | -5.83 | -5.79 |
| A | 232 | 300 | 270 | 114 |
| $W_1$ | 0.42 | 0.40 | 0.40 | 0.40 |
| $W_2$ | 0.23 | 0.20 | 0.21 | - |
| $W_3$ | 0.084 | 0.084 | 0.084 | - |

**Supplementary Table 2** Fittings of conductance histograms using asymmetric peaks (the function is given in Supplementary Methods). Full coverage by FcC$_{11}$SH on 45 nm diameter at +600mV and -600mV, full FcC$_{11}$SH coverage and diluted with C$_{12}$SH (1:1) on 15 nm dots. From plots in Fig.4.

# Supplementary Note 1: History of conductance histograms in molecular electronics

Recognizing that experimental transport through molecules is usually a tunneling process in which the conductance $G$ is defined as $G=Aexp(-\beta d)$ where $A$ is a contact conductance, $\beta$ a tunnel decay constant and $d$ the molecule length, the shape of this peak is often well-described by a log-normal distribution[4-7] (Fig.1b). Peak width (its variance) has been mainly attributed to the adsorption chemistry of the molecule with the electrode, but not discussed quantitatively[5,8-22]. For micrometer-size devices, the log-normal distribution is related to the averaging over various molecular phases and to grain boundaries.

## Supplementary Methods:

## UHV-STM

The high resolution image was performed at room temperature with a substrate biased at 2V and at a constant current of 1 pA. We are not aware of such clear image at molecular (even sub-molecular: see supplementary Fig.1) level on a dense monolayer $FcC_nSH$ monolayer (see ref.[23] for an image obtained with a low density of molecules and ref.[24] for the supramolecular structure obtained when $n<5$). A rectangular crystal-like structure distorted from the ($\sqrt{3}$x$\sqrt{3}$) R30° lattice of Au was previously observed in refs.23, 24. It is rather unexpected for the present monolayer to observe a similar trend as it is believed that Fc-Fc interactions dominate for the short SAMs ($n<5$) and when $n>5$, $C_n$-$C_n$ interactions dominate. We believe that this discrepancy can be attributed to the gentle ultrasonication step in the process. As mentioned in the previous section, this step was required to avoid pollution of the tip from adsorbed molecules on silica. It may bring enough energy to release strain in the monolayer leading to a configuration where the Fc-Fc interactions dominate as for shorter alkyl chains. Interestingly, we find the same tilt angle of Fc (to the normal of the surface) as $FcC_{11}SH$ monolayers with stronger van der Waals interactions. The present crystal-like structure is perfectly adapted to the present study.

### Gold nanodot electrode fabrication

The fabrication of the gold nanocrystal arrays have been described elsewhere.[25] To perform this study, we have developed an high-speed e-beam lithography technique to get large areas covered with Au nanocrystals arrays.[25] For e-beam lithography, an EBPG 5000 Plus from Vistec Lithography was used. The (100) Si substrate (resistivity

= $10^{-3}$ Ω.cm) was cleaned with UV-ozone and the native oxide etched before resist deposition. The e-beam lithography is optimized by using a 45 nm-thick diluted (3:5 with anisole) PMMA (950 K). For the writing, an acceleration voltage of 100 keV was used, which reduces proximity effects around the dots, compared to lower voltages. The dose per dot corresponds to 3-4 fC. The conventional resist development / e-beam Au evaporation (8 nm) / lift-off processes were used. Immediately before evaporation, native oxide is removed with dilute HF solution to allow good electrical contact with the substrate. Single crystal Au nanodots can be obtained after thermal annealing at 260°C during 2 h under $N_2$ atmosphere. During this process, a thin layer of $SiO_2$ covers the dots. This layer is removed by HF at 1% for 1 mn prior to SAM deposition. Minimum spacing between Au nanodots is 50 nm, which is the configuration used for cyclic voltammetry measurements.

### Self-assembled monolayers

For the SAM deposition, we exposed the freshly evaporated gold surfaces and gold nanocrystals to 1 mM solution of 11-ferrocenyl-1-undecanethiol (95% pure from Aldrich) in 80% ethanol (VLSI grade from Carlo Erba) 20% dichloromethane during 24 h in a glovebox in the darkness. Then, we rinsed the treated substrates with ethanol followed by a cleaning with gentle ultrasonication (20% power, 80 kHz) in a bath of chloroform (99% from Carlo Erba) during 1 min. The gentle ultrasonication was required to avoid pollution of the tip from adsorbed molecules on silica. It may play a role in the present molecular organization. In the future, ultrasonication may not be necessary if a monolayer of silane molecules is grafted between dots to prevent non-specific adsorption. For the diluted monolayer of $FcC_{11}SH$ with $C_{12}SH$ molecules (95% pure from Aldrich), both molecules (1:1) were inserted simultaneously, and the grafting/rinsing procedure remained identical. For the gold substrate electrode used for the STM image, 5 nm of Ti and 100 nm of Au were evaporated at 3 Å/s at $10^{-8}$ Torr, the self-assembly process remaining identical to the one for the gold nanodot electrodes.

**Cyclic Voltammetry**

Supplementary Fig. 3 shows the electrochemical cell used in this study. The 0.5 mL container is filled with $NaClO_4$ (0.1M in water) as the electrolyte. The cell is connected with the Solartron ModuLab potentiostat by the three typical electrodes. The nanocrystals operate as the working electrode, a platinum wire as the counter electrode, and a typical Ag/AgCl electrode is used as the reference electrode. Before the experiment, the electrochemical cell is cleaned with ethanol and DI water. "Test" sweeps between -0.1 and 0.6 V with a highly doped silicon substrate (without dots) are measured to confirm that there is no peak due to contamination. Cyclic voltammetry (stable under several voltage cycles) proves the presence of ferrocene-thiol electroactive molecules and allows their quantification. The electrochemical characterization of a sample with few molecules per dot is challenging, in particular in the gold nanoarray structure. Indeed, in Cyclic Voltammetry, the ratio of capacitive current (background noise) to faradic current (signal) is magnified due to the presence of areas (capacitors) between dots without molecules. The Square Wave Voltammetry is the technique of choice to reduce the capacitive contribution. It is based on the fact that the capacitive current decays faster than the Faradic current. The wave has to be optimized to measure the current when the charging current can be considered negligible, but not too low for getting enough signal. Also, due to the increased complexity of the waveform, a calibration procedure has to be done for the SwV (based on cyclic voltammetry) in order to estimate the density of molecules assembled on the dots. A linear regression has been obtained between the CyV at 1V/s and the SwV with the effective scan speed of 0.1 V/s (step: 1 mV, pulse: 20 mV, frequency: 100 Hz, integration period: 100%), the experimental condition selected in this study. CyV and SwV were performed on the same samples with 15 nm diameter dot and at three different dilution ratios (1:0, 1:1 and 1:10).

**Voltammograms fits with consideration of Coulomb interactions**

The extended Laviron model [26] based on Coulomb interactions between redox molecules, leads to the following equations. It enables a direct comparison with experimental data:

$$i(\theta_O) = \frac{n^2 F^2 A v \Gamma_{max}}{RT} \frac{\theta_O (\theta - \theta_O) \theta}{\theta^2 - 2G\theta_O \phi(\theta)(\theta - \theta_O)} \quad \text{(Eq.S1)}$$

where $i$ is the measured voltammetry current, $A$ is the cell area, $\Gamma_{max}$ is the total surface coverage (mol.cm$^{-2}$) of the redox-active species; $n$ is the number of electrons exchange per molecule during a redox process; $v$ is the potential scan rate (V.s$^{-1}$); $F$, $R$ and $T$ are the Faraday constant, the universal gas constant and the absolute temperature, respectively. $\theta_O$ and $\theta_R$ are the oxidized and reduced normalized surface coverage values with $\theta_O + \theta_R = \theta$, $G$ a global constant of interaction, $\Theta$ represents the total fraction of molecules and $\phi(\theta)$ the segregation factor. A versatile function was recently introduced to fit symmetric voltammograms ($I$ vs $E$) with intermolecular interactions as in the present paper[27].

**Estimation of Coulomb interaction energy $\varphi$ from Laviron model and Frumkin isotherm**

Cooperative effects can be discussed quantitatively based on *G* parameter (a Coulomb energy normalized by the thermal energy). *G>0* corresponds to global attractive interactions (voltammogram FWHM below 90 mV) and *G<0* to global repulsive interactions (voltammogram FWHM above 90 mV). The link between *G* and *φ* can be obtained based on the Frumkin isotherm,[28] in the special case of oxidized species interactions:

$$\varphi = -2.k.T.G/N_a \quad (Eq.S2)$$

where $k$ is the Boltzmann constant, $T$ the temperature, and $N_a$ the number of nearest neighbors on the completely covered surface (4 in the present configuration based on the STM image). Values obtained from Eq.S2 correspond to the Coulomb energy interactions between punctual neighboring sites. From fits of Figs.2e-h, we get $G_1 \approx -0.356$ ($G_2 \approx +0.959$) for all dot diameters. This parameter is in the range of previous reports with Fc SAMs[29], although values as large as -1.4 have been reported. Based on Eq.5 with $N_a=4$, it corresponds to $\varphi=4.5$ meV.

**DFT calculations**

The transfer integral has been calculated within the dimer fragment approach, as implemented in the Amsterdam Density Functional package.[30] Dimer orbitals are expressed as a linear combination of molecular orbitals of the fragments, which are obtained by solving Kohn Sham equations. Transfer integrals $t_{12}$ are obtained by computing:

$$t_{ij} = \langle \varphi_i | \hat{H} | \varphi_j \rangle \quad (Eq.S3)$$

where $\varphi_i$ and $\varphi_j$ correspond to HOMO orbitals of the isolated molecules (i.e., fragments). The HOMO is localized over the π orbitals of the cyclopentadienyl rings without a significant weight over the Fe atom. Transfer integral $t_{12}$ has been evaluated at the DFT level[31] by using the B3LYP (Becke, three-parameter, Lee Yang Parr) hybrid functional[32] with a Triple Zeta Polarized basis set. Due to non-orthogonality of the basis set, the transfer integral value is not uniquely defined and depends on the definition of the energy origin[33]. This problem is solved by applying a Löwdin transformation to the initial electronic Hamiltonian, resulting in the following expression of the transfer integral between orthogonal functions used in charge transport models:

$$\tilde{t}_{12} = \frac{t_{12} - (\varepsilon_1 + \varepsilon_2)S_{12}}{1 - S_{12}} \qquad (Eq.S4)$$

where $S_{12}$ represents orbital overlap.

The transfer integral has been calculated for two neighboring Fc units (fragments). Initial position parameters for the neighboring fragments have been deduced from the experimental STM images (base $a$ = 6.778Å, $b$ = 4.675Å, $\gamma$ = 75.78°). The tilt angle formed by Fc and the normal axis is $\theta$ = 60°, in agreement with previous works[Error! Bookmark not defined.] and the STM image. The external electric field barely affects parameter $t$[34].

**Image treatment**

Images are treated on Origin C program. The 1st function (Threshold) applies a threshold to put 0 in the matrix below the threshold, it removes the background noise. Then, 2nd function (maxi) obtains the maximum per dot by checking the nearest neighbors. The functions are available in Supplementary Methods.

**Origin C functions for Histogram generation from CAFM images**

**void Threshold(string strName, double thmin, double thmax, int ibegin, int iend, int jbegin, int jend)**

```
{
	Matrix mm(strName);

	for (int i=ibegin; i<iend; i++)
		for (int j=jbegin; j<jend; j++)
			if ((mm[i][j]<thmin)||(mm[i][j]>thmax)){mm[i][j]=0};
}
```

**void maxi(string strName, int neighbors, int ibegin, int iend, int jbegin, int jend)**

```
{
	Matrix mm(strName);

	for (int i=ibegin; i<iend; i++)
	{
		for (int j=jbegin; j<jend; j++)
		{
			if(mm[i][j]!=0)
			{
				for (int k=-1*neighbors; k<=neighborss; k++)
				{
					for (int l=-1*neighbors; l<=neighbors; l++)
					{
					if ((i+k)>=0)&&((j+l)>=0)&&((i+k)<iend)&&((j+l)<iend))
					if(mm[i+k][j+l]>mm[i][j])
						{mm[i][j]=0};
					}
				}
			}
```

```
            }
    }
    int a=0;
    Worksheet wks;
    wks.Create("histogram.otw");
    WorksheetPage wksp=wks.GetPage();
    wksp.Rename("histogram");

    string str;

    for (int m=ibegin; m<iend; m++)
    {
        for (int n=jbegin; n<jend; n++)
        {
            if (mm[m][n]!=0)
            {
                str.Format("%f",mm[m][n]);
                wks.SetCell(a, 0, str); // set the value to a cell of worksheet
                a++;
            }
        }
    }
}
```

**Theoretical histogram generation (Landauer-Buttiker-Imry formalism)**

Electron transport between two normal electrodes through an array of molecules at temperature $T = 0$ is considered. Each electrode is modeled as a semi-infinite cubic lattice of single-state sites and any site on its surface can be connected to one molecule modeled as a single-level quantum dot (QD).

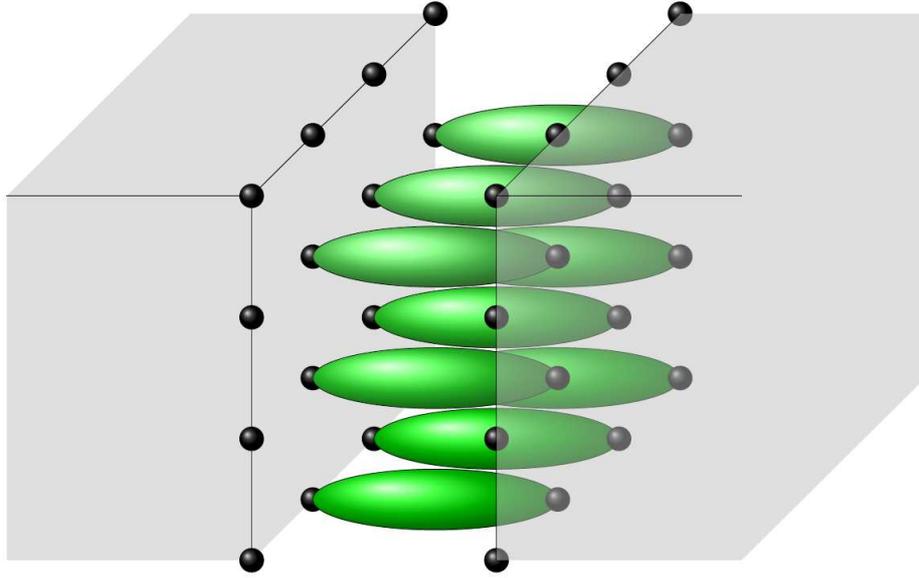

*Two electrodes modeled as semi-infinite 3D tight-binding metals connected via an array of molecules. Each electrode site (black ball) can be connected to one molecule (green ellipse).*

The total Hamiltonian of the system reads:

$$\widehat{H} = \begin{pmatrix} \widehat{H}_{el,R} & \widehat{V}_R^\dagger & 0 \\ \widehat{V}_R & \widehat{H}_{mol} & \widehat{V}_L \\ 0 & \widehat{V}_L^\dagger & \widehat{H}_{el,L} \end{pmatrix} \quad \text{(Eq. S5)}$$

where $\widehat{H}_{mol}$ is the Hamiltonian for the array of molecules and $\widehat{H}_{el,R/L}$ the one of the right and left electrodes, while the couplings between electrodes and molecules read

$$\widehat{V}_{R/L} = \sum_{\alpha,\beta} V_{R/L} |\alpha,\beta\rangle\langle m_\alpha, n_\beta| \quad \text{(Eq. S6)}$$

with electrode-molecule coupling constants $V_R$ and $V_L$ (assumed equal for all molecules). $|\alpha,\beta\rangle$ is the state at site $(\alpha,\beta)$ on the electrode surface while $|m_\alpha, n_\beta\rangle$ is the state of the molecule at position $(m_\alpha, n_\beta)$.

$\hat{H}_{mol}$ can be written generally as:

$$\hat{H}_{mol} = \varepsilon|m_\alpha, n_\beta\rangle\langle m_\alpha, n_\beta| - t[(|m_\alpha, n_\beta\rangle\langle m_\alpha+1, n_\beta| + |\alpha,\beta\rangle\langle m_\alpha, n_\beta+1|) + h.c.]$$

where $\varepsilon$ is each molecule energy level and $t$ is a nearest neighbor transfer integral between molecules, which can be switched on and off. Note that molecules are not only coupled each other, but also via the electrodes: this is a substrate-mediated coupling. These two couplings are collectively termed cooperative effects. Focusing on electronic transport through the molecules, the leads degrees of freedom are integrated out leaving us with the following effective Hamiltonian:

$$\hat{H}_{eff} = \hat{H}_{mol} + \hat{\Sigma}_R(E) + \hat{\Sigma}_L(E) \quad \text{(Eq. S7)}$$

$$\hat{\Sigma}_{R/L}(E) = \hat{V}_{R/L}\,\hat{g}_{el}(E)\hat{V}_{R/L}^\dagger = |V_{R/L}|^2\,\hat{g}_{el}(E) \quad \text{(Eq. S8)}$$

The isolated electrode Green's function $\hat{g}_{el}(E)$ is obtained through Haydock recursion[35], leading to the matrix components

$$g_{|m-m'|,|n-n'|}(E) = \langle m', n'|\hat{g}_{el}(E)|m, n\rangle$$

$$= \frac{1}{2|V_e|^2\pi^2} \int_0^\pi d\theta_1 \cos(|m-m'|\theta_1) \int_0^\pi d\theta_2 \cos(|n-n'|\theta_2)$$

$$\times \Xi\left(\frac{E}{2|V_e|} + \cos\theta_1 + \cos\theta_2\right) \quad \text{(Eq. S9)}$$

$$\Xi(x) = 2x - 2\,sign(x+1)\sqrt{x^2-1} \quad \text{(Eq. S10)}$$

where the nearest-neighbor coupling within the electrodes is set to $V_e = 0.82\ eV$.[4]

It follows that the full Green's function of the molecules dressed by the coupling to the electrodes can be defined as:

$$\hat{G}(E) = [E\hat{I} - \hat{H}_{eff}(E)]^{-1} \quad \text{(Eq. S11)}$$

The transmission through the molecular QDs finally reads:

$$T(E) = Tr\left(\hat{\Gamma}_R(E)\hat{G}(E)\hat{\Gamma}_L(E)\hat{G}^\dagger(E)\right) \quad \text{(Eq. S12)}$$

where we introduced the scattering rate between molecules and electrode $\mu$ as $\hat{\Gamma}_\mu(E) = -2|V_\mu|^2 Im[\hat{g}_{el}(E)]$.

The zero-temperature current through the molecules biased at voltage $V$ is then given by

$$I(V) = \frac{2e}{h}\int_{E_F}^{E_F+eV} dE\ T(E,V) \quad \text{(Eq. S13)}$$

where $E_F$ is the Fermi energy of the electrodes and $T(E,V)$ corresponds to the energy-dependent transmission $T(E)$ where we accounted for a bias-related shift in the dot level energy $\epsilon \to \epsilon(V) = \epsilon + eV\frac{|V_R|^2}{|V_R|^2+|V_L|^2}$.

**Isolated molecule**

The single-molecule Hamiltonian reduces to

$$\hat{H}_{mol} = \epsilon|0,0\rangle\langle 0,0| \quad \text{(Eq.S14)}$$

so that the transmission finally takes the Breit-Wigner-like form

$$T(E,V) = \frac{4|V_R|^2|V_L|^2[Im(g_{00}(E))]^2}{\left|E-\epsilon-eV\frac{|V_R|^2}{|V_R|^2+|V_L|^2}-(|V_R|^2+|V_L|^2)g_{00}(E)\right|^2} \quad \text{(Eq.S15)}$$

Integrating it over the energy window set by the external voltage, this expression allows us to get the current value and check the reasonable choice of coupling parameters.

**Array of $N \times N$ molecules**

We consider a square of $N \times N$ molecules, the site basis being ordered as $\{|1,1\rangle, |1,2\rangle, \dots, |1,N\rangle, |2,1\rangle, |2,2\rangle, \dots, |N,N\rangle\}$. The Hamiltonian of this system

describes the coupling of each molecule to both its horizontal and vertical nearest neighbors, and reads, in the case $N = 3$,

$$\hat{H}_{mol} = \begin{pmatrix} \epsilon & t & 0 & t & 0 & 0 & & & \\ t & \epsilon & t & 0 & t & 0 & & \hat{0} & \\ 0 & t & \epsilon & 0 & 0 & t & & & \\ t & 0 & 0 & \epsilon & t & 0 & t & 0 & 0 \\ 0 & t & 0 & t & \epsilon & t & 0 & t & 0 \\ 0 & 0 & t & 0 & t & \epsilon & 0 & 0 & t \\ & & & t & 0 & 0 & \epsilon & t & 0 \\ & \hat{0} & & 0 & t & 0 & t & \epsilon & t \\ & & & 0 & 0 & t & 0 & t & \epsilon \end{pmatrix} \quad \text{(Eq.S16)}$$

where $\epsilon$ is the molecular energy level and $t$ is the transfer integral, coupling nearest neighbors of the array in both directions.

The electrode Green's function is then expressed in this site basis using Eq. S9 and the horizontal and vertical lengths $|m - m'|$ and $|n - n'|$ between the two considered molecules. In the case of, e.g. $N = 3$, one thus has in this basis

$$\hat{g}_{el}(E) = \begin{pmatrix} g_{00} & g_{01} & g_{02} & g_{10} & g_{11} & g_{12} & g_{20} & g_{21} & g_{22} \\ g_{01} & g_{00} & g_{01} & g_{11} & g_{10} & g_{11} & g_{21} & g_{20} & g_{21} \\ g_{02} & g_{01} & g_{00} & g_{12} & g_{11} & g_{10} & g_{22} & g_{21} & g_{20} \\ g_{10} & g_{11} & g_{12} & g_{00} & g_{01} & g_{02} & g_{10} & g_{11} & g_{12} \\ g_{11} & g_{10} & g_{11} & g_{01} & g_{00} & g_{01} & g_{11} & g_{10} & g_{11} \\ g_{12} & g_{11} & g_{10} & g_{02} & g_{01} & g_{00} & g_{12} & g_{11} & g_{10} \\ g_{20} & g_{21} & g_{22} & g_{10} & g_{11} & g_{12} & g_{00} & g_{01} & g_{02} \\ g_{21} & g_{20} & g_{21} & g_{11} & g_{10} & g_{11} & g_{01} & g_{00} & g_{01} \\ g_{22} & g_{21} & g_{20} & g_{12} & g_{11} & g_{10} & g_{02} & g_{01} & g_{00} \end{pmatrix} \quad \text{(Eq.S17)}$$

Using the symmetry $g_{|m-m'|,|n-n'|}(E) = g_{|n-n'|,|m-m'|}(E)$ one can reduce this $N^2 \times N^2$ matrix down to only $\frac{N(N+1)}{2}$ independent coefficients. Once $\hat{g}_{el}$ is known, the self-energy $\hat{\Sigma}$ entering the effective Hamiltonian can be computed, ultimately giving access to the full Green's function $\hat{G}(E)$ expressed in the same site basis, and finally to the energy-dependent transmission $T(E)$.

**Computing transport histograms**

In order to compare to the experimental results, we would like to compute current histograms. Each histogram typically relies on $10^6$ realizations which requires a very large number of integrals (Eq.S13), making the computation untractable. This is particularly true for large values of *N*. In order to speed up the numerical computations, we need to significantly reduce the number of evaluations of the transmission. To do so, we approximate the current integral using a midpoint rule:

$$I(V) = \frac{2e}{h} \, eV \, T\left(E_F + \frac{eV}{2}, V\right). \qquad (Eq.S18)$$

As it is the case for any Newton-Cotes-like scheme, the error is related to the second derivative of the integrand. Focusing on an applied bias *V*=-0.6 V, one readily sees that the transmission, although not quite linear, is monotonous and only weakly convex over the domain of integration (Fig.S14a), therefore justifying our approximation. For this choice of external bias, the current is thus dominated by the transmission evaluated at the center of the voltage window. Fig.S14b shows the ratio of the approximated current (Eq.S18) over the current obtained from the integrated transmission (Eq.S13) as a function of the applied voltage. The error from the approximation made with Eq.S18 is acceptable at the studied bias given that the current distribution spreads over more than one decade. We stress that a larger error arises at positive bias. In any case, further studies would benefit from a refined evaluation of the current integral (using e.g. Simpson's rule) concomitantly with an overall optimization of the numerical computations.

At fixed voltage $V$, and for a given array of $N \times N$ molecules, transmission histograms are evaluated assuming a Gaussian distribution of all four relevant parameters of the effective Hamiltonian $\widehat{H}_{eff}$, thus requiring a total of **8** parameters to be determined, namely

$\epsilon$: *molecular energy*

$\delta\epsilon$: *molecular energy standard deviation*

$V_{t/b}$: *electrode-molecule coupling constants*

$\delta V$ : *standard deviation of the coupling constants*

$t$: *average inter-molecule coupling energy*

$\delta t$: *standard deviation of the coupling energy between molecules*

**Estimating the relevant parameters**

Focusing on the particular case of an external voltage bias $V = -0.6\ eV$, we could determine the values of the coupling constants $V_{t/b}$ by computing the current in the isolated molecule case (numerically integrating Eq. S15 over energy) and matching it to the experimentally measured value.

Moreover, experimental data suggest that the molecular energy level $\epsilon$, measured from the Fermi energy of the electrodes is of the order of $0.2\ eV$. From electrochemistry results, we could infer that the standard deviation of the molecular energy is bounded, so that $\delta\epsilon < 0.045\ eV$. In order to probe the effect of this distribution in molecular energy, we typically consider several possible values of $\delta\epsilon$ within this interval. However, when studying the influence of other parameters, we assume an average value of $\delta\epsilon = 40\ meV$.

We now focus on the experimental situation of (1:1) dilution. This situation corresponds to weak cooperative effects and is more accurately described in our case by the isolated molecule approach. Assuming for simplicity that $\delta V_t = \delta V_b$, this allows us to determine the value of $\delta V$ upon generating the conductance histogram which best reproduces the experimentally measured log-normal shape.

We then move on to the (1:0) dilution ratio. There, the cooperative effects are expected to be strong and we use our model relying on an array of $N \times N$ molecules. We thus need to determine the values of the two remaining parameters, namely the average inter-molecule coupling energy $t$ and the corresponding standard deviation $\delta t$. These are obtained from the current histogram that best approaches the asymmetric double sigmoidal function (Eq.4) where the values of the various parameters are presented in supplementary Table 2.

In order to perform a systematic determination of these parameters, we introduce a new quantity $d_{fit}$ which measures the distance in the space of real-valued functions between the asymmetric function $f(x)$ that fits the experimental data, and the transmission histogram generated from the present theoretical approach

$$d_{fit} = \sum_{i=1}^{N_{bin}} \frac{|H_i - f(x_i)|}{N_{bin}} \qquad (Eq.S19)$$

where $i$ labels the $N_{bin}$ bins of the histogram, of height $H_i$ and centered on $x_i$. We then determine the values of $t$ and $\delta t$ that minimize this distance $d_{fit}$ to the fitting function.

**Normalizing histograms**

In order to compare experimental and theoretical results, we need to find a common normalization as different histograms can be constructed with different number of total counts. One possibility is to scale the experimental data so that the tallest bin corresponds to a count of 1. This in turn leads to a fitting function equivalent to the one in Eq. S18, only with a new parameter $A'$ (in place of $A$), accounting for this rescaling. It follows that the total area covered by the fitting function is now

$$S = \int_{-\infty}^{\infty} dx \; A' \frac{1}{1+e^{-(x-x_c+w_1/2)/w_2}} \left(1 - \frac{1}{1+e^{-(x-x_c-w_1/2)/w_3}}\right) \approx 0.436 \, A' \quad \text{(Eq.S20)}$$

where we used the values of $w_1$, $w_2$ and $w_3$ from supplementary Table 3. We then apply a rescaling prefactor to the transmission histograms generated according to the formalism described above in order for their total area to precisely match the one in Eq. S20.

# Supplementary References: